\documentclass[useAMS,usenatbib]{mnras}
\usepackage{graphicx}
\usepackage{multirow}
\usepackage{lipsum}
\usepackage{blindtext}
\usepackage{amsmath}
\usepackage{amssymb}

\newcommand{\ra}[1]{\renewcommand{\arraystretch}{#1}}

\title[The mass-size relation of LRGs from BOSS $\&$ DECaLS]{The mass-size relation of LRGs from BOSS and DECaLS}

\author[Favole et al. 2018]
{\parbox[t]{\textwidth}{\vspace{-0.7cm}Ginevra Favole$^{1}$\thanks{E-mail: gfavole@sciops.esa.int}, Antonio D. Montero-Dorta$^2$, Francisco Prada$^3$, Sergio A. Rodr\'iguez-Torres$^{4,5,6}$, David J. Schlegel$^7$}
\vspace*{15pt}\\ 
$^1$European Space Astronomy Centre (ESAC), 28692 Villanueva de la Ca\~nada, Madrid, Spain\\
$^2$Instituto de F\'isica, Universidade de S\~ao Paulo, S\~ao Paulo, SP, Brazil\\
$^3$Instituto de Astrof\'isica de Andaluc\'ia (IAA)/ CSIC, Granada, E-18008, Spain\\
$^4$Instituto de F\'{i}sica Te\'{o}rica (IFT) UAM/CSIC, Universidad Aut\'{o}noma de Madrid, Cantoblanco, E-28049 Madrid, Spain\\
$^5$Campus of International Excellence UAM/CSIC, Cantoblanco, E-28049 Madrid, Spain\\
$^6$Departamento de F\'isica Te\'orica M-8, Universidad Aut\'onoma de Madrid, Cantoblanco, 28049 Madrid, Spain\\
$^7$Lawrence Berkeley National Laboratory, 1 Cyclotron Road, Berkeley, CA, 94720, USA\\
\vspace{-0.8cm}
}
\date{  }

\begin{document}
\pagerange{\pageref{firstpage}--\pageref{lastpage}} \pubyear{2018}
\maketitle

\begin{abstract}
We use the DECaLS DR3 survey photometry matched to the SDSS-III/BOSS DR12 spectroscopic catalog to investigate the morphology and stellar mass-size relation of luminous red galaxies (LRGs) within the CMASS and LOWZ galaxy samples in the redshift range $0.2<z<0.7$. The large majority of both samples is composed of early-type galaxies with De Vaucouleurs profiles, while only less than 20\% are late-type exponentials. We calibrate DECaLS effective radii using the  higher resolution CFHT/MegaCam observations and optimise the correction for each morphological type. 
By cross-matching the photometric properties of the early-type population with the Portsmouth stellar mass catalog, we are able to explore the high-mass end of the distribution using a large sample of 313,026 galaxies over 4380\,deg$^{2}$. 
We find a clear correlation between the sizes and the stellar masses of these galaxies, which appears flatter than previous estimates at lower masses.
The sizes of these early-type galaxies do not exhibit significant evolution within the BOSS redshift range, but a slightly declining redshift trend is found when these results are combined with $z\sim0.1$ SDSS measurements at the high-mass end. The synergy between BOSS and DECaLS has important applications in other fields, including galaxy clustering and weak lensing.  
\end{abstract}



\section{Introduction}
\label{sec:intro}

The SDSS-III/Baryon Oscillation Spectroscopic Survey \citep[BOSS;][]{Eisenstein2011, Dawson2013} provided unprecedented statistics at the high-mass end by measuring the spectra of about 1.5 million luminous red galaxies \citep[LRGs\,;][]{Eisenstein2001} over 10,000\,deg$^2$ of sky down to magnitude  $r\sim 22.2$ and within the redshift range $0.2<z<0.7$. This data set has been used not only to accurately measure the baryon acoustic oscillation feature \citep[BAO;][]{Eisenstein2005, Anderson2014, 2017MNRAS.470.2617A}, but also to study the massive galaxy population at $z\sim0.55$. BOSS allowed us to characterise the red/blue color bimodality observed in LRGs \citep[]{2013MNRAS.432..359T, 2014MNRAS.437.1109R, 2016MNRAS.462.2218F, AMD2016a}, to constrain the high-mass end of the stellar mass  and luminosity functions of these massive galaxies \citep[][]{2013MNRAS.435.2764M, 2013MNRAS.436..697B, 2016MNRAS.457.4021L, 2016MNRAS.455.4122B, 2017MNRAS.467.2217B, AMD2016a} and to measure the intrinsic relation between galaxy luminosity and velocity dispersion \citep[]{AMD2016b, 2017MNRAS.468...47M}. Despite these achievements, the morphological and structural properties of BOSS LRGs have been difficult to probe due to the poor SDSS image quality (median seeing of 2").

More recently, the Dark Energy Camera Legacy Survey\footnote{\url{http://legacysurvey.org/decamls/}} (DECaLS) of the SDSS Equatorial Sky has been designed to obtain high-quality images that cover 6700$\,\rm{deg^2}$ in three optical bands $(g,\,r,\,z)$. With a limiting magnitude of $r\leq 23.4$ and a median seeing of $1.2"$, it allows a narrower and more efficient target selection for the DESI survey \citep[][]{2013MNRAS.428.1498C, 2016A&A...592A.121C}. DECaLS improves dramatically the quality of the SDSS data set, providing also deeper photometry. 

Besides the classification of galaxies through their morphology and shape parameters, the stellar mass-size relation has been explored in a number of works as a powerful scaling law to connect fundamental galaxy properties. \citet[]{2010MNRAS.404.2087B} studied the distribution of stellar mass (M$_{\star}$), size, velocity dispersion, luminosity and color as a function of galaxy morphology and concentration index for SDSS massive early-type galaxies. They claimed that sample selections based on colour or concentration lead to significantly different scaling relations. \citet[]{2011MNRAS.412L...6B} investigated further these dependencies in a sample of SDSS early-type galaxies (ETGs) and found that there is a particular stellar mass scale (M$_{\star}\sim2\times10^{11}\,\rm{M_{\odot}}$) beyond which major mergers start to dominate the assembly histories of these massive galaxies. \citet[]{2013ApJ...778L...2C} identified the same mass scale as the transition point between two processes that regulate the mass-size distribution of galaxies in dense environments and in the field. From one side, spiral galaxies are replaced by bulge-dominated fast-rotator ETGs, with the same mass-size relation and mass distribution as in the field. On the other hand, the slow-rotator ETGs are segregated in mass from the fast ones, and their size increases proportionally to their mass. 
These evidences suggest that bulge growth (outside-in evolution) and bulge-related environmental quenching dominate in the low-mass end, while dry mergers (inside-out evolution) and halo-related quenching shape the mass and size growth at the high-mass end. 

\citet[]{2013ApJ...779...29H} investigated the impact of different large-scale environments  (i.e., field, group and clusters) on the size of massive ETGs at $z\sim0$. At fixed stellar mass, they did not find any significant dependence of the central and satellite ETG sizes on the environment. The mass-size relation of these galaxies is independent of the host halo mass and the galaxy position within the halo. This result is not sensitive to different galaxy selections based on morphology, star formation, or central density. 
\citet[]{2011MNRAS.415.3903T} studied the buildup of the mass-size relation of elliptical galaxies from $z\sim0$ up to $z\sim1$, using observations from SDSS and HST/GOODS. They did not find any evidence for age segregation at fixed stellar mass. This rules out the scenario of a present-day mass-size relation progressively established through a bottom-up sequence in which older galaxies populate its lower tail, remaining in place since their formation. Their result supports instead the hypothesis that the local mass-size relation is defined at $z\sim1$, with all galaxies occupying a region half of the size of the present-day distribution.
\citet[]{2003MNRAS.343..978S} explored the connection between galaxy size and luminosity (or stellar mass) using $z\sim0.1$ SDSS data and found a trend which is significantly steeper for early- than for late-type galaxies. 

Recently, \citet[]{2017arXiv170704979Z} analysed the dependence of the luminosity- or mass-size relation on galaxy concentration and morphology in the SDSS DR7 Main galaxy sample. They found a clear trend of smaller sizes and steeper slope for early-type elliptical galaxies.
\citet[][]{2011MNRAS.418.1055M} studied the morphology and size of BOSS luminous massive galaxies using HST/COSMOS photometry and found that about $74\%$ of them are early-type elliptical or lenticular, while the rest are late-type spirals. \citet[]{2014ApJ...789...92B} compared galaxy size measurements in SDSS, SDSS-III/BOSS and COSMOS data at $0.1\lesssim z\lesssim 0.7$ to derive accurate corrections for the galaxy effective radii (i.e. sizes). \citet[][]{2017ApJ...837..147H} investigated the redshift-size relation in massive ETGs in the UltraVISTA and CANDELS surveys. They found evidence of a significant mass build up at $r<3\,$kpc beyond $z>4$, and a clear evolutionary change at $z\sim1.5$, when the galaxy progenitor stops growing in-situ through disk star formation and  accretes minor mergers. \citet[][]{2017arXiv170103526S} explored the ratio between galaxy size and dark matter halo virial radius at $z\lesssim3$ using data from GAMA and CANDELS. They found very little dependence on stellar mass and lower ratios at high redshift for more massive galaxies.  

In this work, we aim to characterise the morphology and the stellar mass-size relation of the well-known SDSS-III/BOSS DR12 CMASS and LOWZ galaxy samples \citep[]{Anderson2012, Bolton2012, Anderson2014, 2015ApJS..219...12A} within the redshift range $0.2<z<0.7$. To this purpose, we match these BOSS spectroscopic samples to the DECaLS DR3 photometric catalog. We calibrate DECaLS sizes using the high-resolution (0.6" median seeing) CFHT/MegaCam observations and optimise the correction individually for each morphological type.
By cross-matching our DECaLS selections with the Portsmouth \citep[][]{2013MNRAS.435.2764M} stellar mass catalog at $0.2<z<0.7$, we are able to constrain the M$_{\star}$--size relation of very massive LRGs in a sample of unprecedented size at these redshifts. Our cross-matched BOSS-DECaLS galaxy samples with CFHT calibrated sizes are made publicly available for the community on the \textsc{Skies and Universes}\footnote{\url{http://www.skiesanduniverses.org/}} database.

The paper is organised as follows: Section \ref{sec:data} describes the data sets used in our analysis. In Section \ref{sec:calibration} we explain how the DECaLS effective radii are calibrated using CFHT observations. In Section \ref{sec:results} we present our results: the morphology of BOSS galaxies, their stellar mass-size relation and their size evolution. We compare with previous studies in Section\,\ref{sec:comparison} and summarize our conclusions in Section \ref{sec:summary}. For the analysis we adopt the cosmology: $h=0.6777,\,\Omega_m=0.3071,\,\Omega_{\Lambda}=0.6929,\,n=0.96,\,\sigma_8=0.8228$ \citep[][]{Planck2014}.


\section{Data and galaxy selections}
\label{sec:data}

\begin{figure}
\begin{center}
\includegraphics[width=0.96\linewidth]{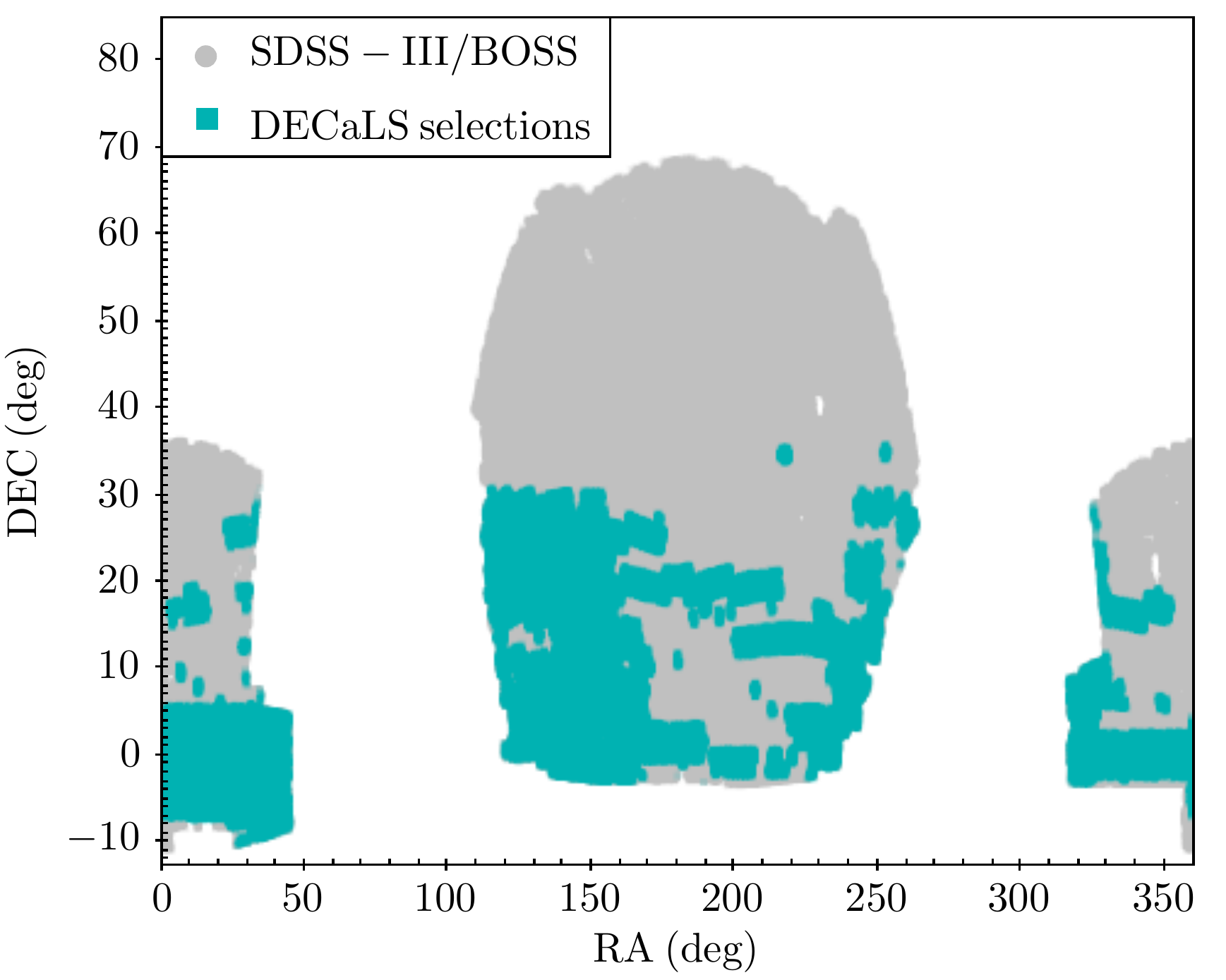}\hfill
\caption{Footprint of the cross-matched DECaLS-BOSS galaxy sample (green area) versus the original SDSS-III/BOSS coverage (grey).}
\label{fig:completeness}
\end{center}
\end{figure}
\begin{figure}
\begin{center}
\includegraphics[width=1.04\linewidth]{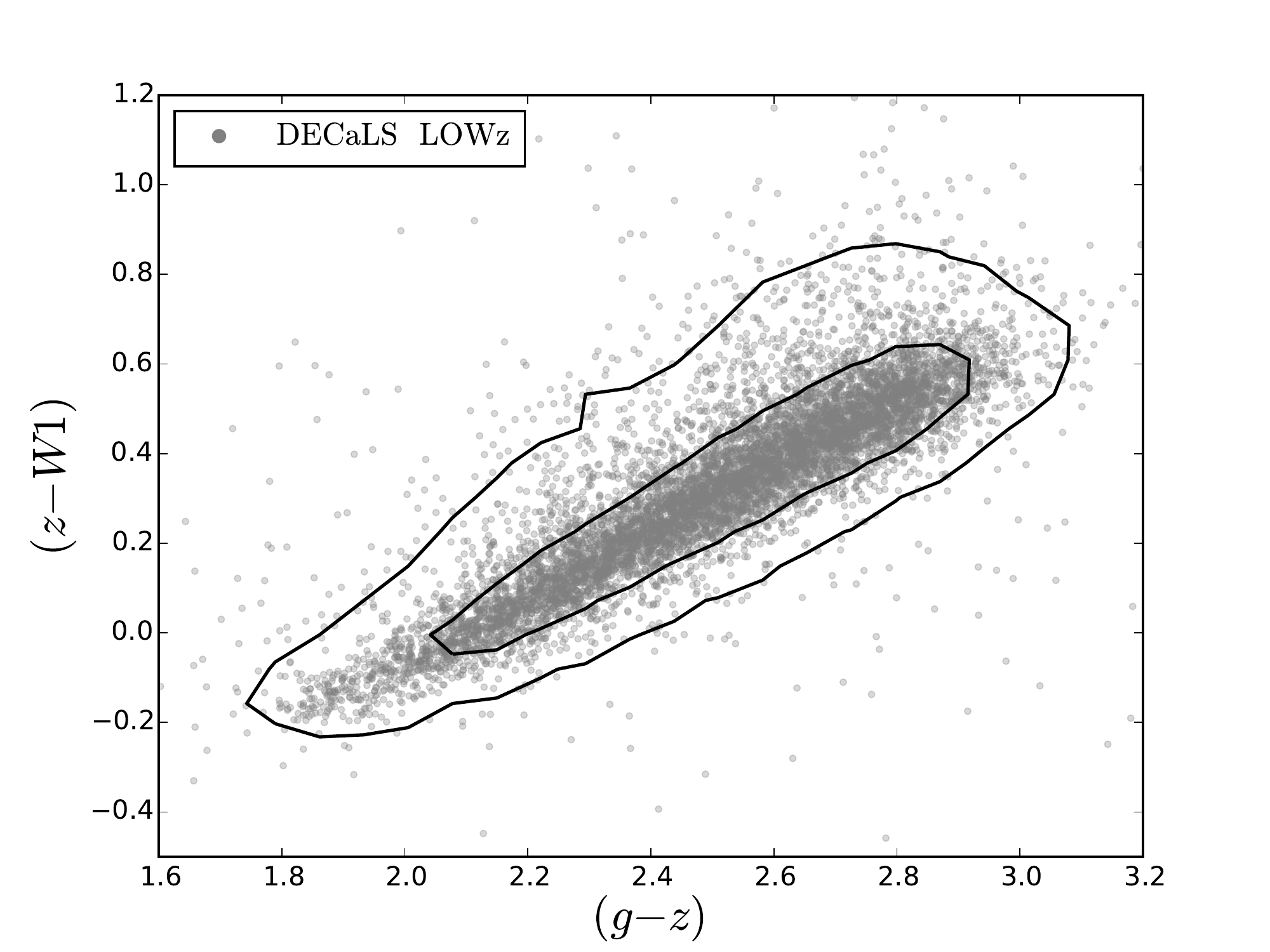}\hfill\vspace{-0.4cm}
\includegraphics[width=1.04\linewidth]{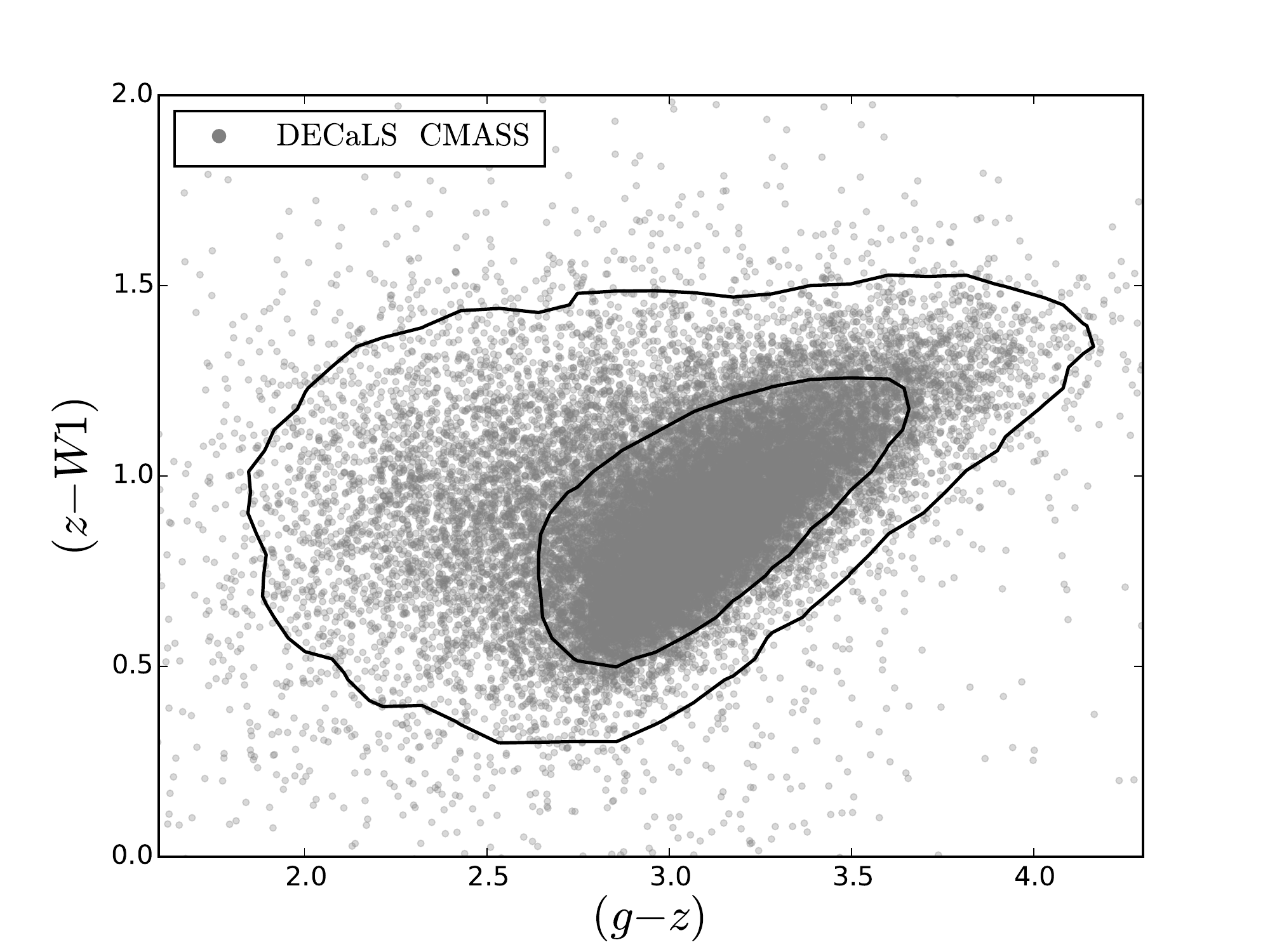}\hfill
\caption{$(z-W1)$ vs. $(g-z)$ color distributions of the cross-matched DECaLS-BOSS LOWZ (top) and CMASS (bottom) samples. The contours denote the $1\,\sigma$ and $2\,\sigma$ uncertainty regions.}
\label{fig:selezioni}
\end{center}
\end{figure}

We use the DECam Legacy Survey (DECaLS) DR3 photometric catalog\footnote{\url{http://legacysurvey.org/dr3/files/}} row-by-row-matched to the SDSS DR12 spectroscopic galaxy sample\footnote{\url{https://data.sdss.org/sas/dr12/sdss/spectro/redux/}}. DECaLS is an optical survey on the 4m Blanco telescope at Cerro Tololo Inter-American Observatory designed to complement the SDSS, SDSS-III, SDSS-IV and DESI surveys with high-quality images from 6700$\,\rm{deg^2}$ of extragalactic sky in the equatorial region in three optical bands $(g,\,r,\,z)$. The DECaLS DR3 photometric catalog also includes the infrared WISE\footnote{\url{http://wise.ssl.berkeley.edu/index.html}} bands (W1, W2, W3, W4). The sky coverage lies within $-18^{\circ} < \delta < +34^{\circ}$ in celestial and $|b| > 18^{\circ}$ in Galactic coordinates. DECaLS has improved dramatically the quality of the SDSS imaging data, providing a deeper photometry with limiting magnitude of $r\leq 23.4$ and a median seeing of $1.2"$.

In the cross-matched catalog introduced above, we select the BOSS CMASS and LOWZ galaxy samples of LRGs (hereafter our \textquoteleft\textquoteleft parent samples") using the SDSS spectroscopic flags\footnote{\url{http://www.sdss.org/dr13/algorithms/boss_galaxy_ts/}}.

\noindent We further exclude point-like sources from the parent samples by imposing the DECaLS condition \texttt{TYPE!="PSF"}. We recover 238,008 CMASS and 75,018 LOWZ galaxies, respectively, i.e., about 31\% and 23\% of the original BOSS samples. The missing galaxies are not observed by DECaLS DR3, which has an effective area of 4380\,deg$^{2}$, much smaller than the 9376\,deg$^{2}$ of the SDSS-III/BOSS, as shown in Figure \ref{fig:completeness}. In Figure \ref{fig:selezioni}, we display our LOWZ (top panel) and CMASS (bottom) parent samples in the DECaLS color-color plane. We use the $g$ and $z$-band magnitudes from DECaLS and the W1 infrared magnitude from WISE to highlight the color properties of BOSS LRGs in DECaLS photometry.

Beside DECaLS magnitudes, for the analysis we adopt DECaLS effective radii, surface brightness profiles and galaxy morphologies. We perform galaxy size calibrations using data from two different surveys: the MegaPrime/MegaCam\footnote{\url{http://www.cfht.hawaii.edu/Instruments/Imaging/MegaPrime/}} at CFHT and the Cosmic Evolution Survey (COSMOS)\footnote{\url{http://cosmos.astro.caltech.edu/}}. The first one has a $1\,$deg$^2$ field-of-view with a resolution of 0.187" per pixel and a median seeing of $\sim 0.7"$. It provides much better imaging quality, which is key to precisely determine galaxy sizes and morphological types. The second survey was originally designed to probe galaxy formation and evolution over a 2\,deg$^2$ equatorial field with imaging by most of the major space-based telescopes and a number of large ground based telescopes.

We adopt \citet[][]{2013MNRAS.435.2764M} stellar masses for the galaxies in our parent samples to study the mass-size relation of LRGs at $0.2<z<0.7$. These are estimated by fitting model spectral energy distributions to the BOSS observed magnitudes. 


\section{Galaxy size calibration}
\label{sec:calibration}

\begin{figure*}
\begin{center}
\includegraphics[width=0.48\linewidth]{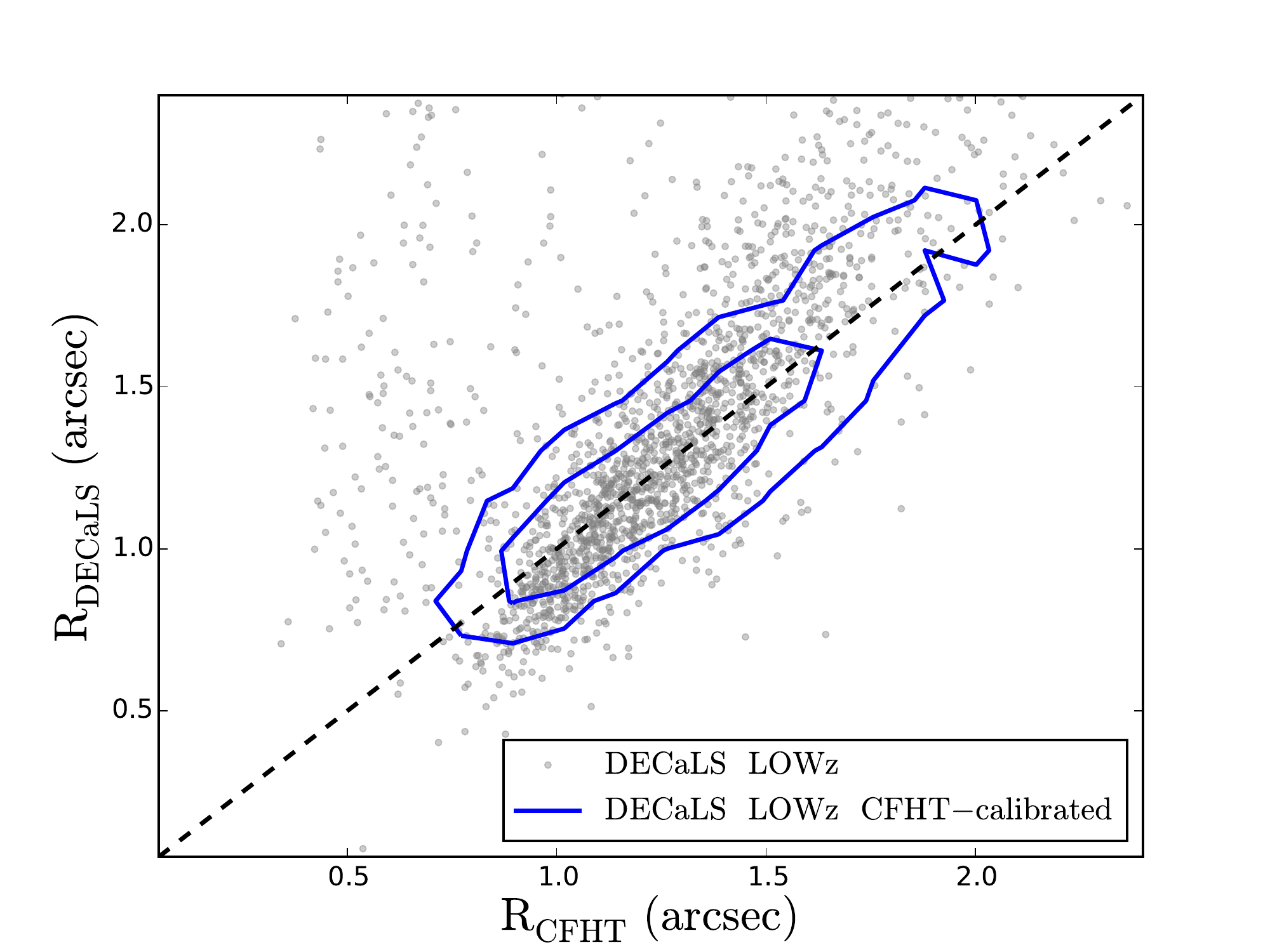}\hfill
\includegraphics[width=0.48\linewidth]{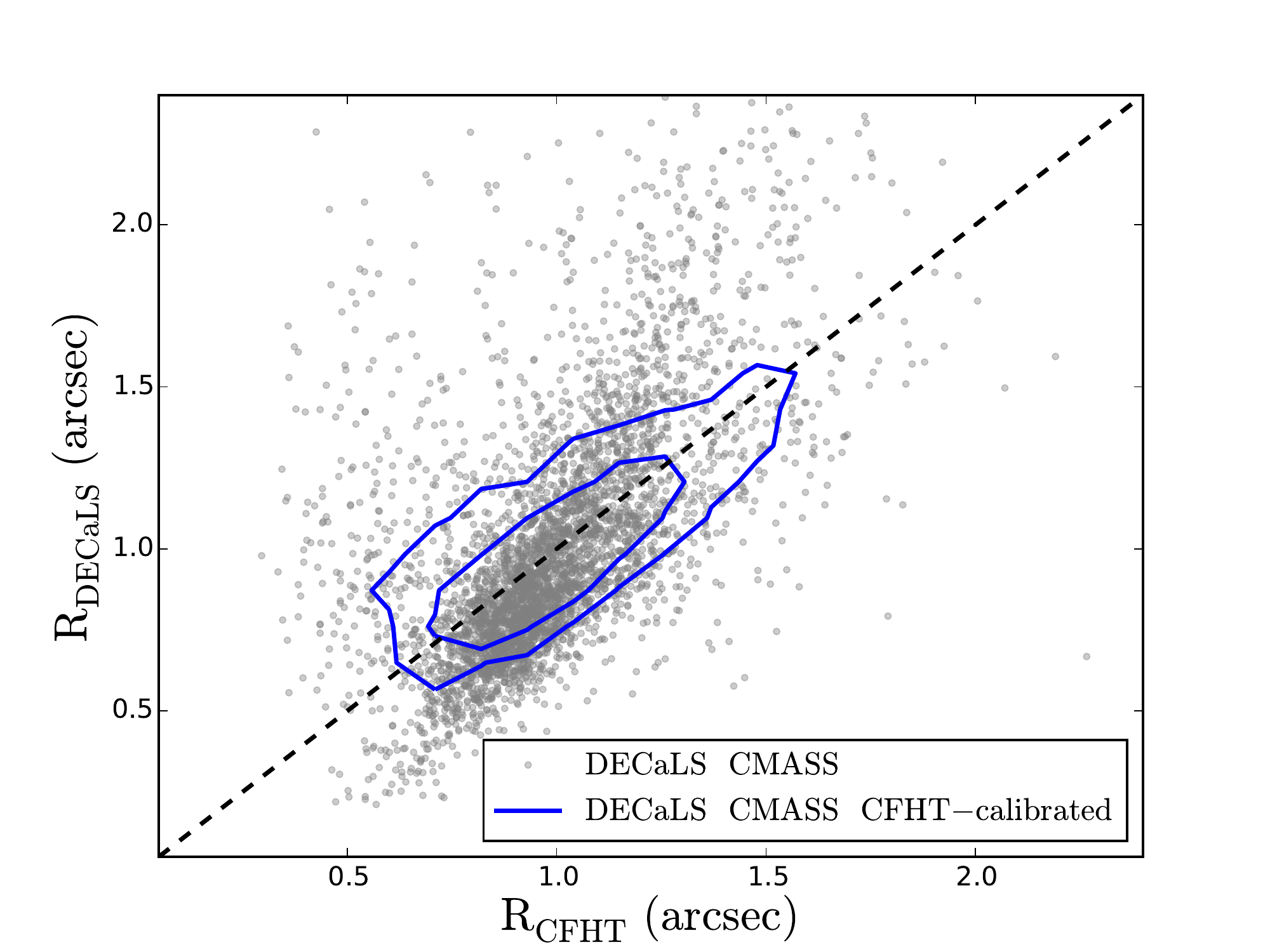}\hfill
\includegraphics[width=0.48\linewidth]{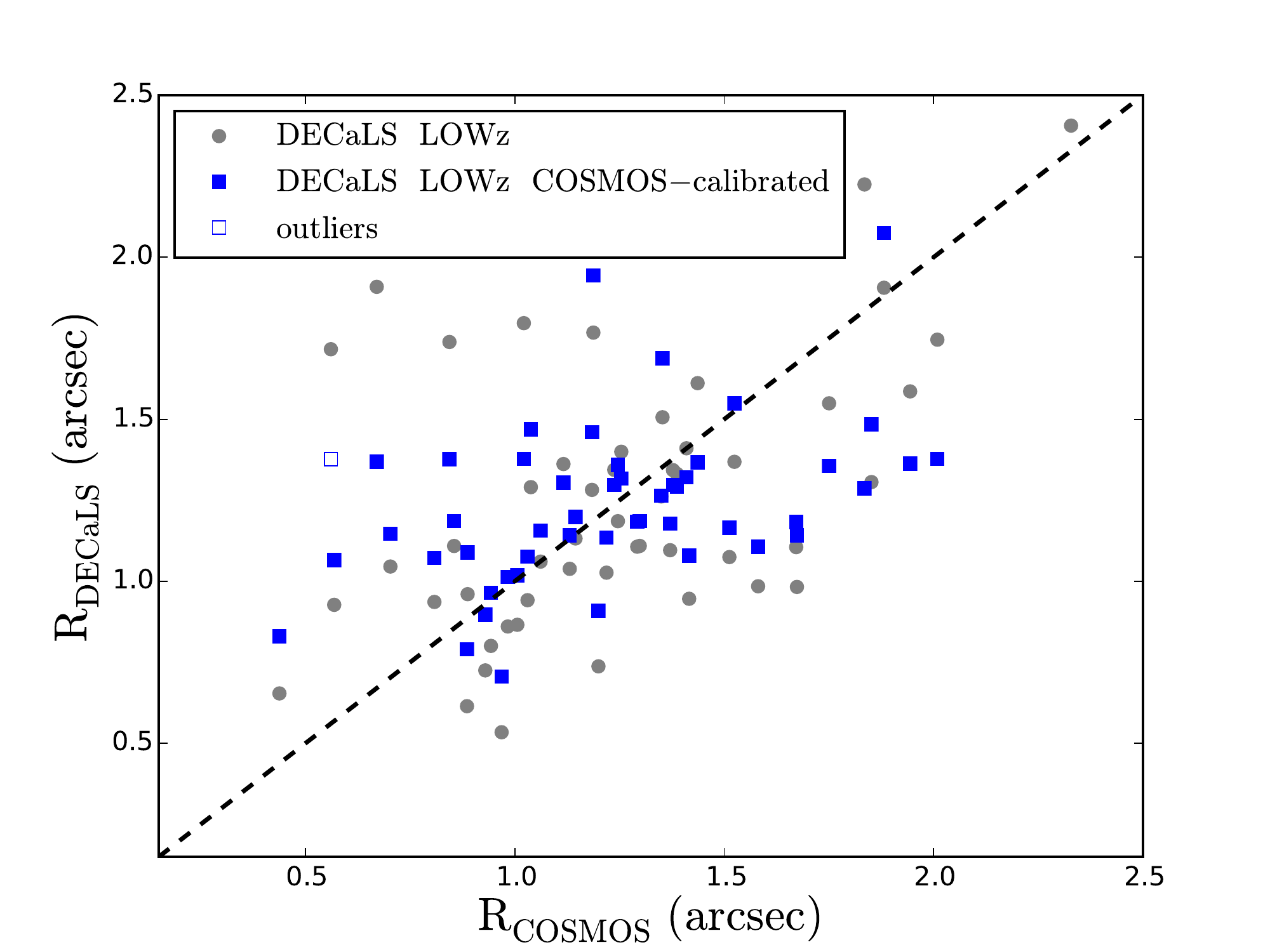}\hspace{0.6cm}
\includegraphics[width=0.48\linewidth]{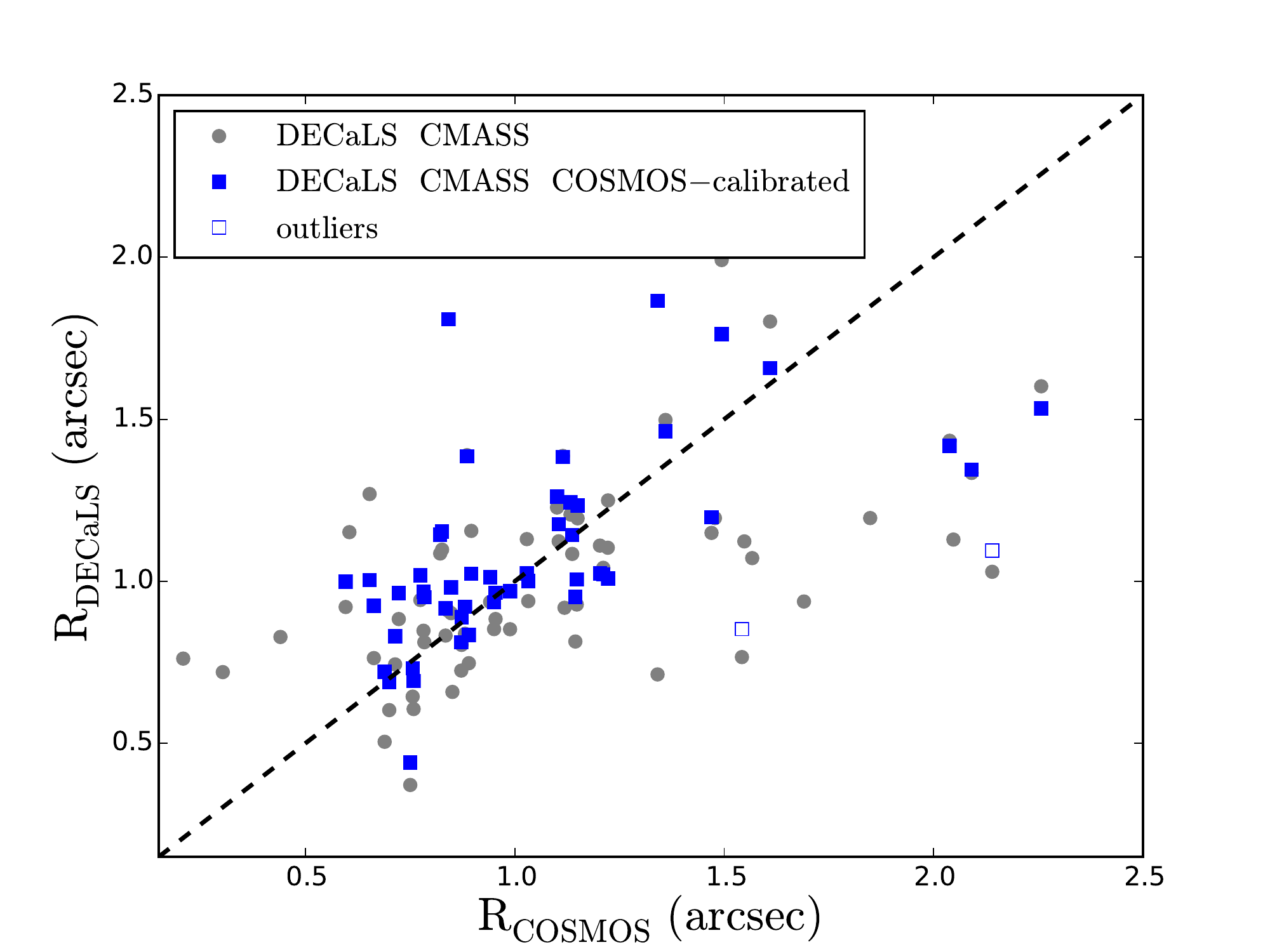}\hfill 
\caption{DECaLS LOWZ (left column) and CMASS (right column) effective radii as a function of the corresponding CFHT (top row) and COSMOS (bottom row) sizes in arcsec. The dashed diagonal line corresponds to the 1:1 relation for each case.}
\label{fig:corr}
\end{center}
\end{figure*}

\begin{table*}\centering
\ra{1.3}
\begin{tabular}{@{}lccccc@{}}   
\hline
&  \multicolumn{2}{c}{DECaLS LOWZ} &   &\\
\hline
 &$0.15\leq z<0.3$         & $0.3\leq z<0.43$ &   &  \\
 & $R_0$\,[arcsec]\hspace{1cm}$\alpha$   & $R_0$\,[arcsec]\hspace{1cm}$\alpha$ & & \\
\hline
CFHT DeV & \,1.226$\pm$0.020\hspace{0.5cm} -0.324$\pm$0.014 &1.141$\pm$0.009 \hspace{0.5cm}-0.395$\pm$0.011 &  & \\
CFHT Exp &\, 1.370$\pm$0.168\hspace{0.5cm} -0.672$\pm$0.154 &1.292$\pm$0.094 \hspace{0.5cm}-0.652$\pm$0.143 &  &\\
\hline
COSMOS DeV & \,1.241$\pm$0.289 \hspace{0.5cm}-0.079$\pm$0.166&1.556$\pm$0.168\hspace{0.5cm}   -0.439$\pm$0.128&     & \\
COSMOS Exp  &  \hspace{0.5cm}-- &\hspace{0.5cm}-- &     &\\
\hline
\hline
&   \multicolumn{2}{c}{DECaLS CMASS}   &&\\
\hline
 &    $0.43< z\leq 0.55$&$0.55< z<0.7$ && \\
 &   $R_0$\,[arcsec]\hspace{1cm}$\alpha$ &$R_0$\,[arcsec] \hspace{1cm} $\alpha$ && \\
\hline
CFHT DeV &   \,           1.009$\pm$0.006\hspace{0.5cm} -0.469$\pm$0.009 &0.952$\pm$0.006 \hspace{0.5cm}-0.547$\pm$0.011&& \\
CFHT Exp &    \,             2.085$\pm$0.147\hspace{0.5cm}-0.276$\pm$0.020 &2.123$\pm$0.143 \hspace{0.5cm}-0.247$\pm$0.018 &&\\
\hline
COSMOS DeV & \,                1.256 $\pm$0.126\hspace{0.5cm}-0.186$\pm$0.083 &1.847$\pm$0.356 \hspace{0.5cm}-0.832$\pm$0.344&& \\
COSMOS Exp  &  \hspace{0.5cm}-- &\hspace{0.5cm}-- &    &\\
\hline
\end{tabular}\vspace{0.3cm}
   \caption{Best-fit coefficients for the calibration factor $f(R_{\rm{DECaLS}})$ given in Eq. \ref{eq:calib}. The COSMOS correction for the DECaLS CMASS and LOWZ samples with exponential profile is omitted due to the lack of statistics.}
 \label{tab:corrtable}
 \end{table*}
 
In order to correct our galaxy size measurements from seeing effects \citep[][]{1993MNRAS.264..961S,2003AJ....125.1882B,2014ApJ...789...92B}, we calibrate DECaLS effective radii with the latest CFHT (see Section \ref{sec:data}) observations. We cross-match our CMASS and LOWZ samples with the data available in the four CFHT fields. Only galaxies with De Vaucouleurs and exponential profiles are employed. For those objects surviving the matching (4721 in CMASS and 2050 in LOWZ), we compare their radii measured in both surveys. We define the DECaLS circularized radius as $R_{\rm{DECaLS}}=R_{\rm{eff}}\sqrt{(b/a)}$, where $R_{\rm{eff}}$ is the DECaLS effective radius, while $a$ and $b$ are the semi-major and semi-minor ellipse axes, respectively. For the calibration we use the following functional form:
\begin{equation}
R_{\rm{DECaLS}}^{\rm{calib}}=R_{\rm{DECaLS}}\times f(R_{\rm{DECaLS}}),
\label{eq:correctionfunc}
\end{equation}
where $f(R_{\rm{DECaLS}})$ is the calibration function depending on DECaLS size defined as:
\begin{equation}
f(R_{\rm{DECaLS}})=\left (\frac{R_{\rm{DECaLS}}}{R_0}\right )^\alpha.
\label{eq:calib}
\end{equation}
We separately fit CMASS and LOWZ galaxies with De Vaucouleurs and exponential profiles to find the optimal parameters $\alpha$ and $R_0$. As part of the fitting procedure, we perform sigma-clipping, rejecting those objects located more than $2\,\sigma$ away from the mean of the $R_{\rm{CFHT}}/R_{\rm{DECaLS}}$ distribution. The excluded points are considered outliers in what follows. The best-fit parameters are reported in Table \ref{tab:corrtable}.
In the top panels of Figure \ref{fig:corr}, we display DECaLS versus CFHT effective radii of the LOWZ (left) and CMASS (right) samples, respectively. The grey points are DECaLS original radii before the CFHT calibration; the blue contours are the corrected sizes. 
The effect of the CFHT calibration lowers DECaLS effective radii by a $\sim$40\% factor, fully consistent with the statistical correction made by \citet[][]{2011MNRAS.418.1055M} using the Zurich Estimator of Structural Types \citep[ZEST;][]{2007ApJS..172..406S} measurements. In what follows, we extrapolate and apply this calibration to the entire CMASS and LOWZ parent samples.
 
In order to test the CFHT calibration, we also derive an independent correction by cross-matching DECaLS with COSMOS data. Even though the overlap between the two data sets is very small -- only 67 galaxies survive the matching for CMASS and 56 for LOWZ -- the result is consistent with the CFHT analysis, as shown in the bottom panels of Figure \ref{fig:corr}. Here we show DECaLS LOWZ on the left and CMASS on the right side. The grey points are the DECaLS radii before correction and the blue filled squares are the sizes calibrated using COSMOS data. The blue empty squares are the outliers, i.e. those objects located more than $2\,\sigma$ away from the mean of the corrected distribution.


\section{Results}
\label{sec:results}
In this section, we present our main results: the morphology of the cross-matched BOSS-DECaLS CMASS and LOWZ samples and the stellar mass-size relation for their early-type galaxy population.
\subsection{The morphology of BOSS LRGs}
\label{sec:morfologia}
\begin{figure*}
\begin{center}
\hspace{-0.5cm}
\includegraphics[width=0.45\linewidth]{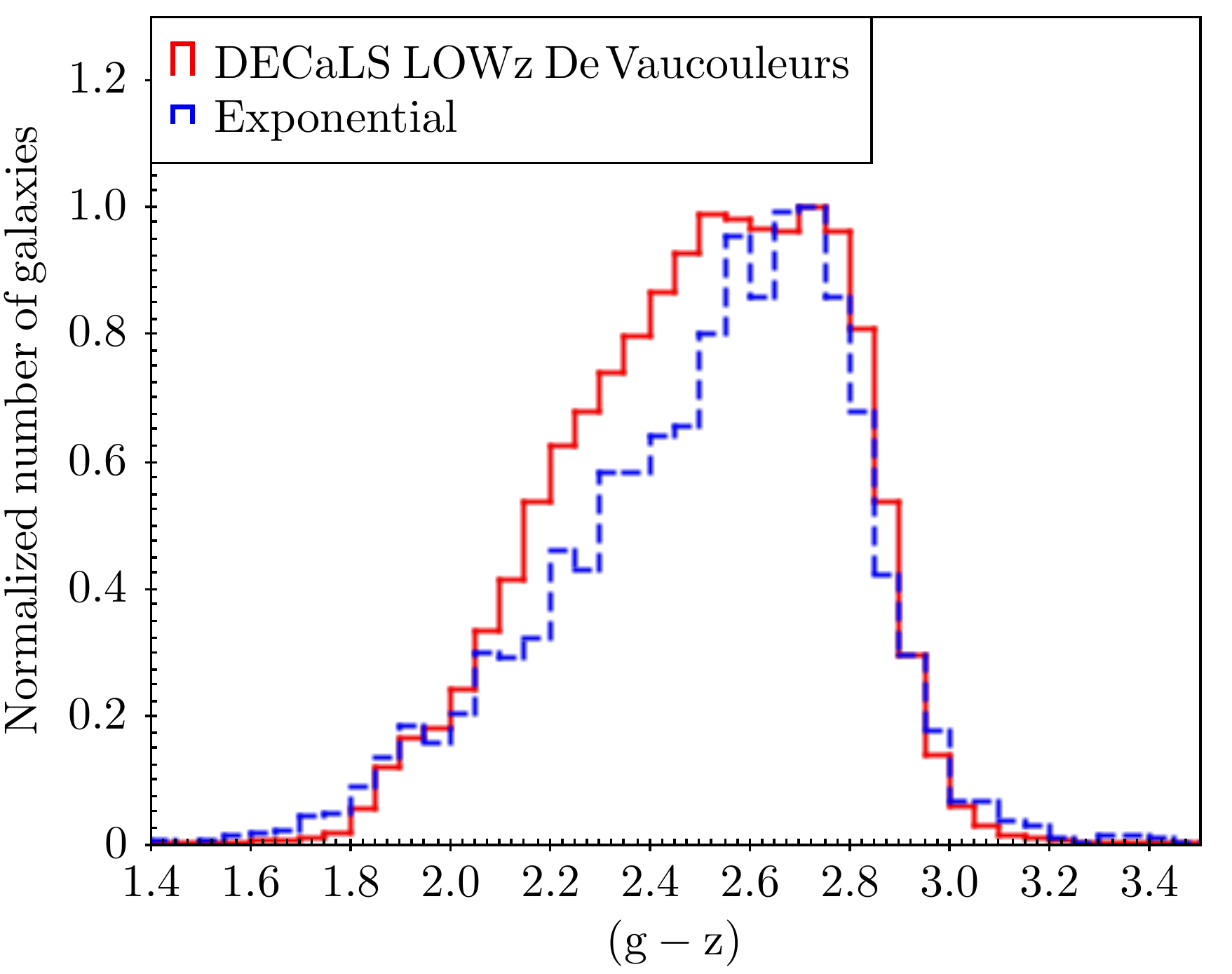}\hfill\hspace{0.4cm}
\includegraphics[width=0.45\linewidth]{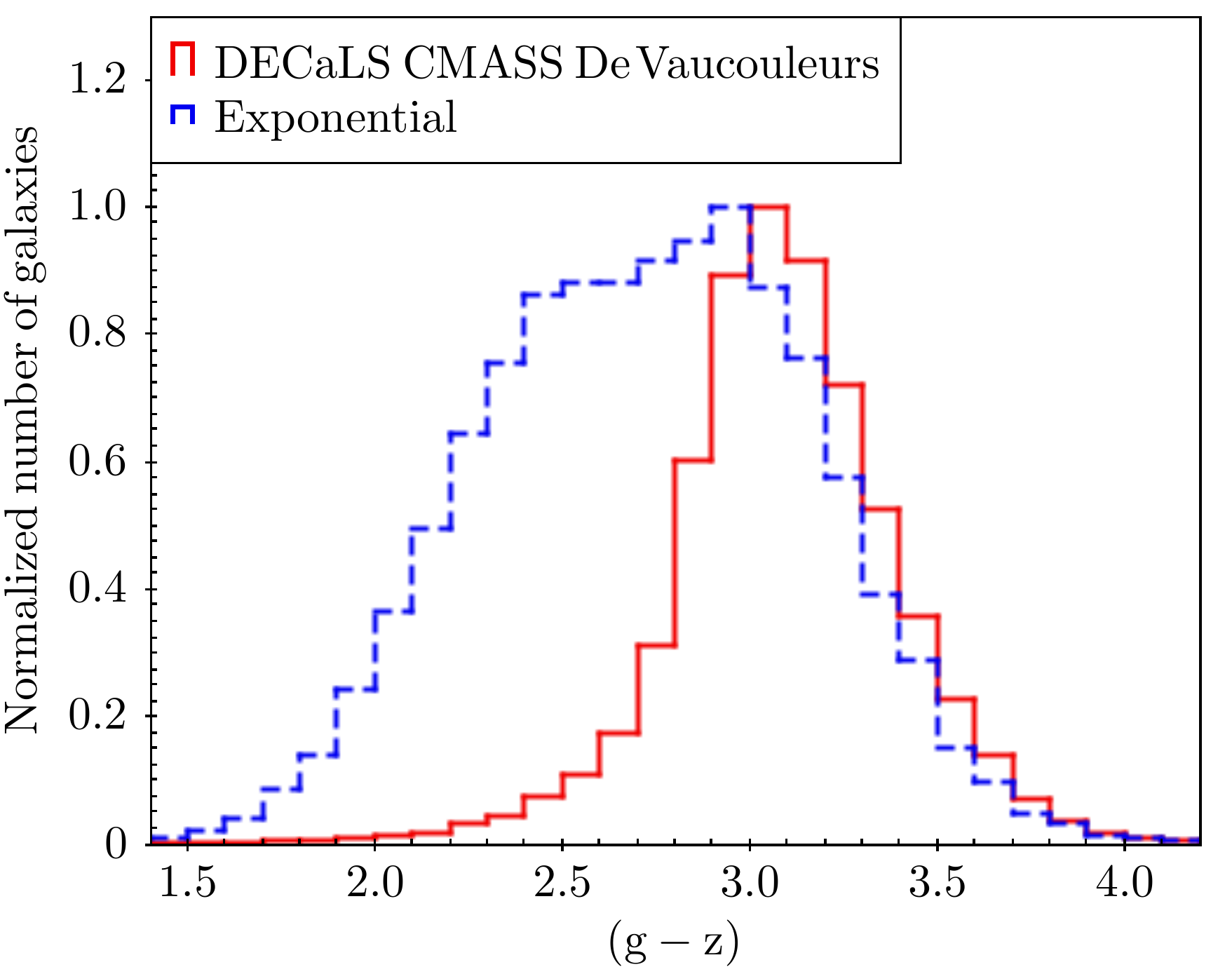}\hfill
\caption{$(g-z)$ color distribution of the cross-matched DECaLS-BOSS LOWZ (left) and CMASS (right) samples. The contributions of galaxies with De Vaucouleurs and exponential profiles are shown as red, solid and blue dashed histograms, respectively.}
\label{fig:giplot}
\end{center}
\end{figure*}

\begin{figure*}
\begin{center}
\includegraphics[width=0.5\linewidth]{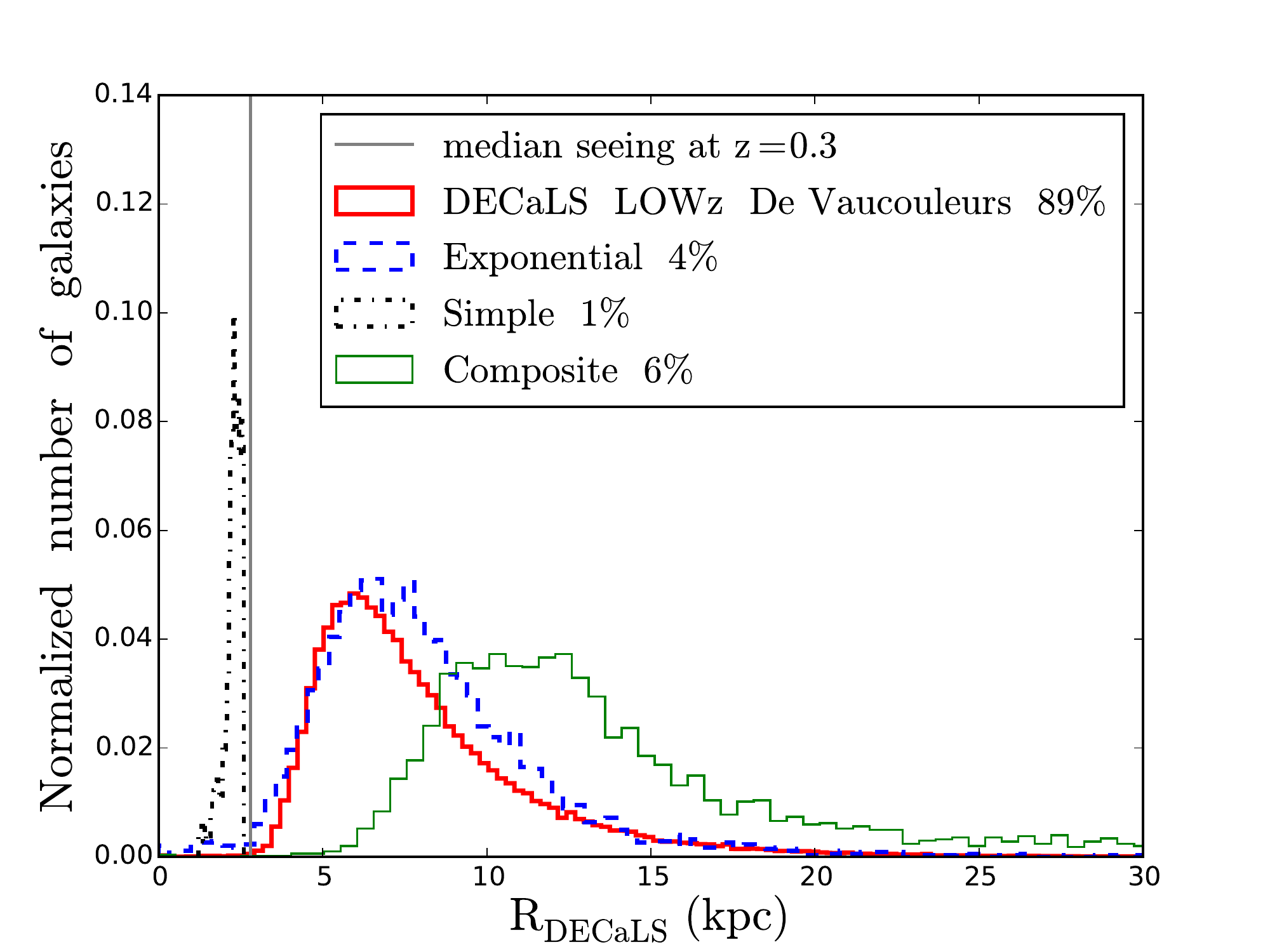}\hfill
\includegraphics[width=0.5\linewidth]{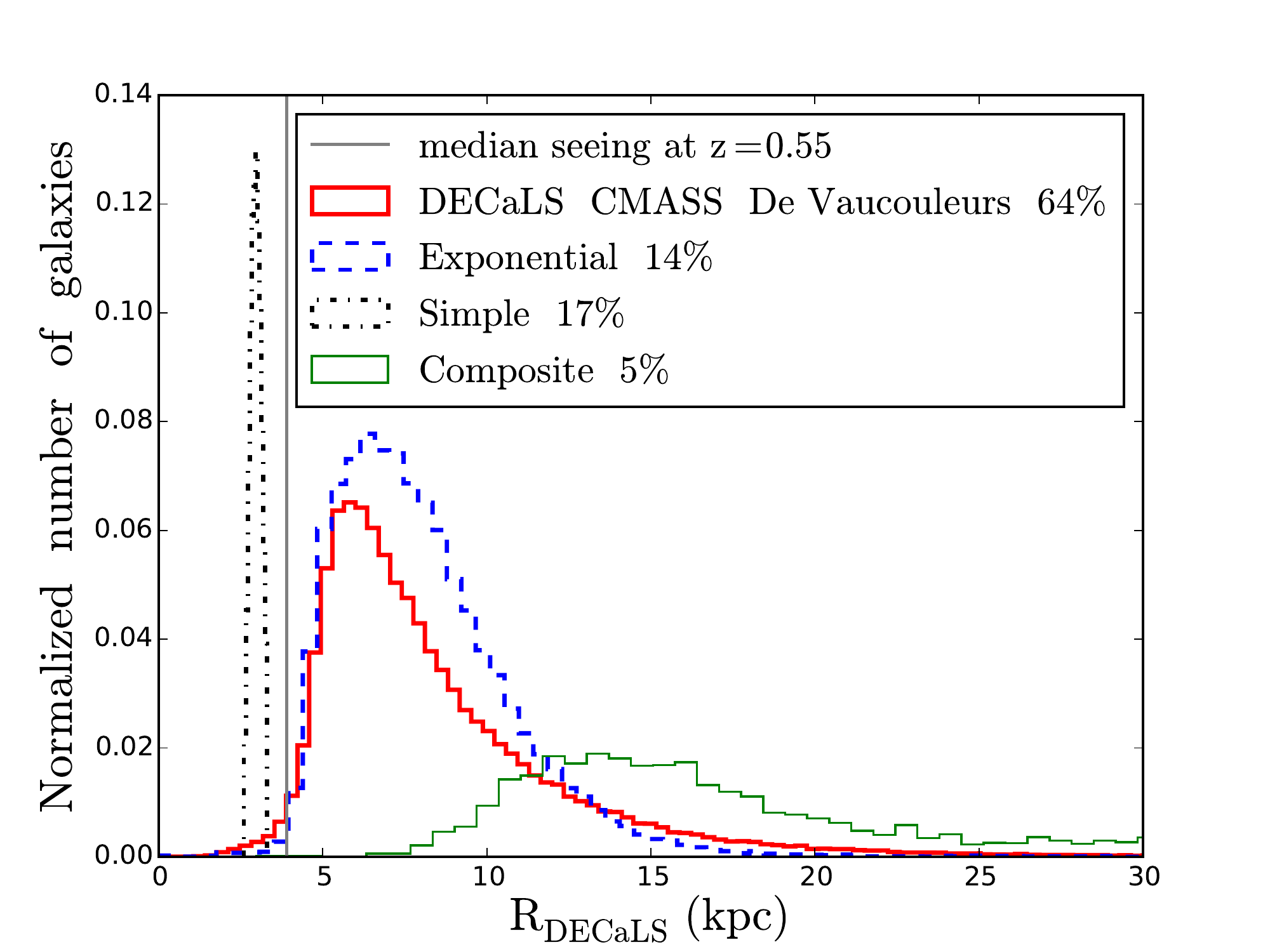}\hfill
\caption{DECaLS LOWZ (left) and CMASS (right) effective radius. The large majority (89\% in LOWZ and 64\% in CMASS) of both samples is composed by galaxies with De Vaucouleurs profiles. Only 4\% (14\%) of LOWZ (CMASS) galaxies in DECaLS have an exponential profile. Objects classified as \textquoteleft\textquoteleft simple" have exponential profiles and round shape, with fixed effective radius. Galaxies classified as  \textquoteleft\textquoteleft composite" are fitted by a combination of De Vaucouleurs and exponential profiles. The vertical dashed lines represent the median DECaLS seeing at the mean redshift of each sample.}
\label{fig:hlr}
\end{center}
\end{figure*}

We use the DECaLS surface brightness profile classification as an indicator of the morphology of CMASS and LOWZ galaxies. In DECaLS, the following profiles have been fitted to individual objects: 
\begin{itemize}
\item De Vaucouleurs: Sersic \citep[][]{1968adga.book.....S} profile with $n=4$.
\item Exponential: Sersic profile with $n=1$. 
\item Composite: linear combination of a De Vaucouleurs and an exponential profile with the same source center.  
\item Simple: exponential profile with a fixed 0.45" effective radius and circular shape. 
\end{itemize}
We find that 64\% (89\%) of CMASS (LOWZ) galaxies have De Vaucouleurs profiles; 14\% (4\%) are exponentials; 17\% (1\%) are simple and 5\% (6\%) are composite. 
Galaxies with De Vaucouleurs profiles are typically early-type/ellipticals, while exponentials correspond to late-type/spirals \cite[see e.g.,][]{1993MNRAS.265.1013C, 1994MNRAS.271..523D, 1995MNRAS.275..874A, 2005AJ....129...61B, 2007ApJ...659.1159S, 2011A&A...529A..53T}. 
Composite profiles are a mixture of the two previous configurations. Simple profiles are used when any other profile with varying radius does not yield a significantly better $\chi^2$ (note that the number of parameters is penalized in the determination of the goodness of fit). 

The CMASS selection allows for a fraction of bluer objects in the sample, which increases with redshift \citep[]{Eisenstein2001, AMD2016a}. This explains the presence of galaxies with exponential profiles. Interestingly, the fraction of galaxies with De Vaucouleurs profiles increases significantly from the LOWZ to the CMASS sample, as the fraction of exponentials decreases. In Figure \ref{fig:giplot}, we show the 
$(g-z)$ color distributions in both the LOWZ (left) and CMASS (right) samples for De Vaucouleurs and elliptical galaxies separately. In the CMASS sample, galaxies with De Vaucouleurs profiles are significantly redder than those showing an exponential profile, as expected from the early-late type association. Interestingly, this separation is less obvious in the LOWZ sample, which might be due to the presence of more dusty spirals having an exponential profile. Note that the red/blue separation in the CMASS sample is more evident in the $(g-i)$ color distribution (i.e., $(g-i)=2.35$), as shown in \citet[][]{2011MNRAS.418.1055M}, \citet[][]{Dawson2013}, \citet[][]{2013MNRAS.435.2764M}, \citet[][]{2014MNRAS.437.1109R}, \citet[][]{2016MNRAS.462.2218F}, and \citet[][]{2017ApJ...836...87L}. 

The fraction of late-type and early-type galaxies that we find in our samples is approximately consistent, given the uncertainties and differences between different methods, with results from \citet[]{2011MNRAS.418.1055M}, \citet[]{2013MNRAS.435.2764M} and \citet[]{AMD2016a} using the SDSS photometry.

In Figure \ref{fig:hlr}, we show the effective radius distribution of the LOWZ (left) and CMASS (right) samples, highlighting the contribution from the different morphologies.  In both populations, the early-type De Vaucouleurs galaxy distribution peaks at $\rm{R_{DECaLS}}\sim7\,\rm{kpc}$, exponentials around $8\,\rm{kpc}$, composite at $12\,\rm{kpc}$ and simple below $5\,\rm{kpc}$. Most of the galaxies classified as ``composite" have a companion nearby preventing to accurately measure their effective radius. Due to this configuration, composite galaxies have on average larger radii and wider size distributions compared to the other morphologies. The number of galaxies and the number density (per unit deg$^2$) of each sample are reported in Table \ref{tab:densities}.

\begin{table*}\centering
\ra{1.3}
\begin{tabular}{@{}lccccrrrc@{}}
\hline
& \multicolumn{3}{c}{DECaLS LOWZ} & \phantom{abc}& \multicolumn{3}{c}{DECaLS CMASS} &
\phantom{abc} \\
\hline
& N$_{\rm{gal}}$ & $n_{\rm{dens}}$\,[deg$^{-2}$] & fraction\,[\%] && N$_{\rm{gal}}$ &$n_{\rm{dens}}$\,[deg$^{-2}$] & fraction\,[\%]  \\
\hline
Total & 84,986 & 19.4 & 100 &&  239,431 & 54.7 & 100\\
De\,Vaucouleurs & 75,441 & 17.2 & 89  && 154,004 & 35.2 & 64\\
Exponential & 3464& 0.8& 4 &&33,681 & 7.7 & 14 \\
Simple & 1062& 0.3& 1 &&41,292 & 9.4 & 17 \\
Composite &5019& 1.1& 6  && 10,454 & 2.4 & 5\\
\hline
\end{tabular}\vspace{0.3cm}
   \caption{The number, number density (per unit deg$^2$) and fraction of De Vaucouleurs, exponential, simple and composite galaxies in the DECaLS LOWZ and CMASS samples.}
 \label{tab:densities}
 \end{table*}
In Figure \ref{fig:hlr}, the median seeing at the corresponding redshift of each sample is represented by a solid vertical line. The DECaLS PSF is dominated by seeing on scales of 1-1.2", which corresponds to a FWHM of about 2.8\,kpc at the mean redshift of LOWZ ($z\sim0.3$) and about 3.9\,kpc at the mean redshift of CMASS ($z\sim0.55$). This makes the effective radius distribution fall sharply at small radii. For LOWZ galaxies, however, this effect is less pronounced due to their larger angular size compared to CMASS objects. In what follows, we exclude from our samples those objects classified as ``simple", which have effective radius significantly lower than these thresholds. 

\subsection{The mass-size relation of LRGs at $0.2<z<0.7$}
\label{sec:Mstar}
Hereafter, we will focus only on LRGs with De Vaucouleurs profiles. 
\begin{figure*}
\begin{center}
\includegraphics[width=0.48\linewidth]{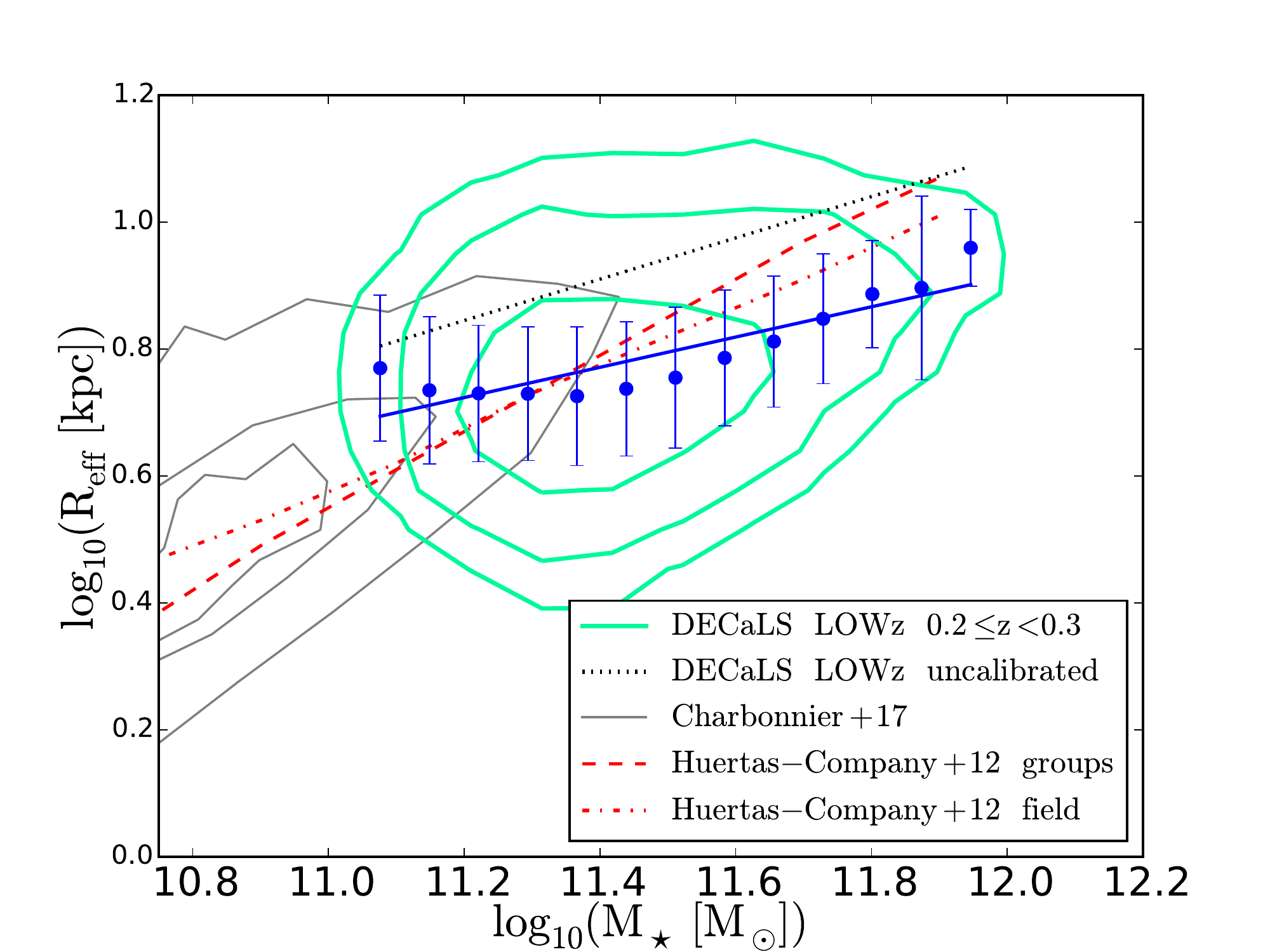}\hfill 
\includegraphics[width=0.48\linewidth]{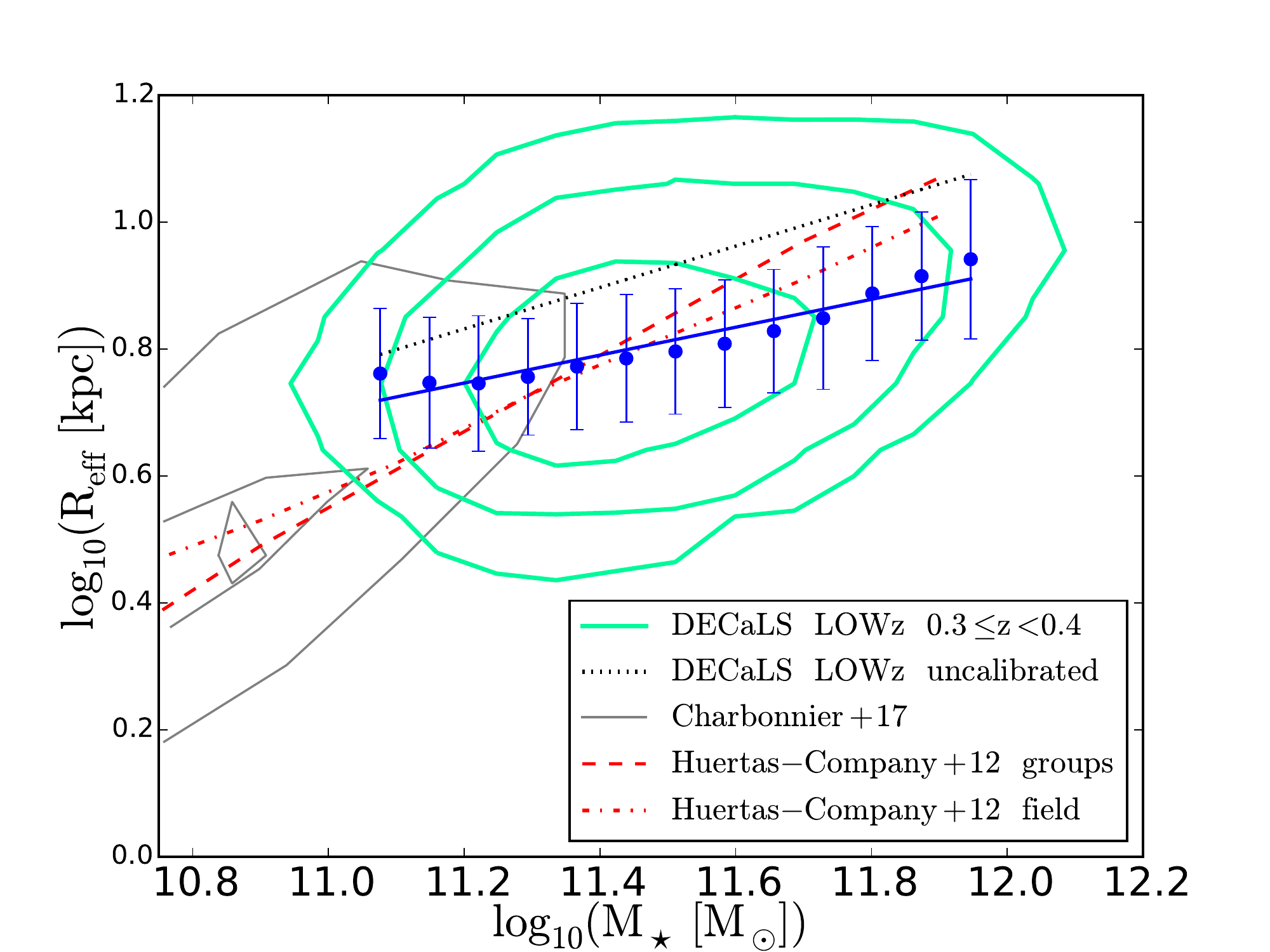} \hfill
\includegraphics[width=0.48\linewidth]{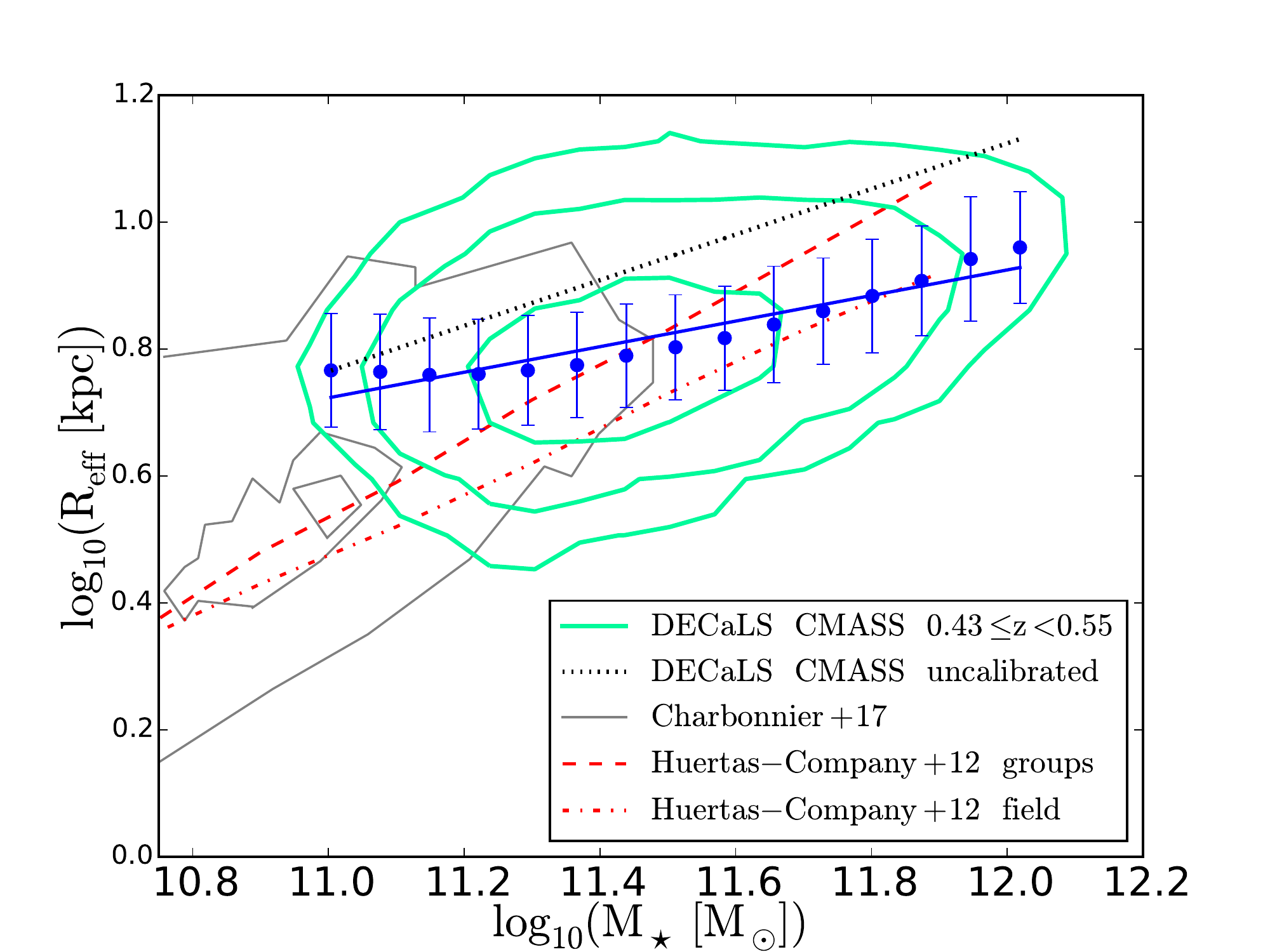}\hspace{0.6cm}
\includegraphics[width=0.48\linewidth]{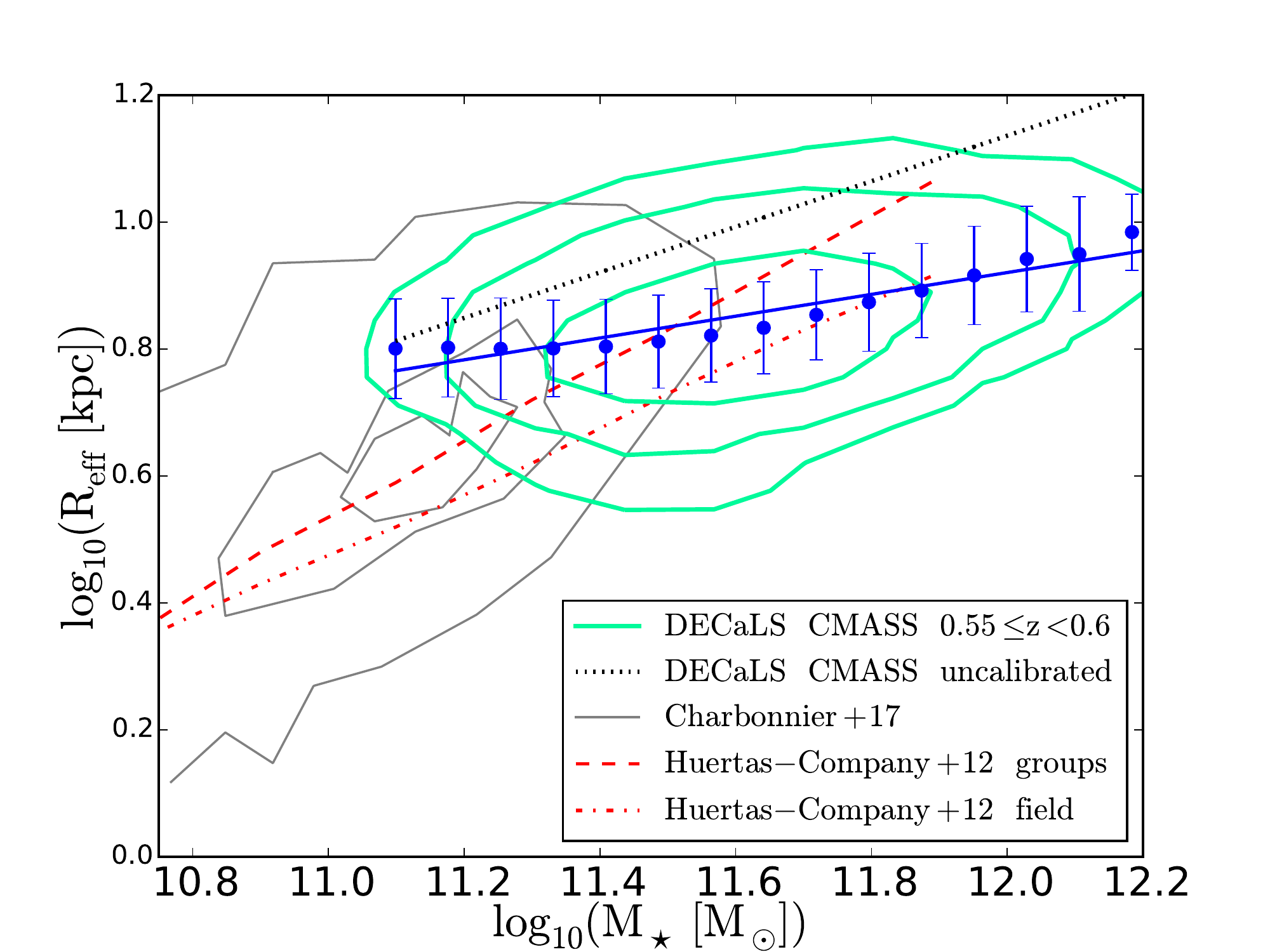}
\caption{Stellar mass--\,size relation for the DECaLS LOWZ (top row) and CMASS (bottom row) samples, considering only galaxies with De Vaucouleurs profiles. DECaLS effective radii are calibrated using CFHT data as explained in Section \ref{sec:calibration}. We show in green the $1\sigma$ (innermost), $2\sigma$ (median) and $3\sigma$ (outermost) contours of each distribution, weighted against stellar mass incompleteness by applying the correction from \citet[]{2016MNRAS.457.4021L}. The blue points are the mean radii in bins of stellar mass and the error bars are the $\pm1\,\sigma$ scatter. The blue solid line is a linear fit to these mean values. The black dotted line is the linear fit to the uncalibrated relation. The grey thin contours correspond to previous observations of less massive quiescent galaxies in CFHT SDSS Stripe 82 \citep[][]{2017MNRAS.469.4523C}. The red dashed and dot-dashed lines are the results for COSMOS ETGs in groups and in the field environment from \citet[]{2013MNRAS.428.1715H}.}
\label{fig:masssize}
\end{center}
\end{figure*}
\begin{table*}\centering
\ra{1.3}
\begin{tabular}{@{}lccccc@{}}
\hline
& \multicolumn{2}{c}{DECaLS LOWZ} &  \multicolumn{2}{c}{DECaLS CMASS} &  \\
\hline
&   $0.2\leq z<0.3$ & $0.3\leq z<0.4$ &    $0.43\leq z< 0.55$&$0.55\leq z<0.6$  \\
\hline
$f$&1.00 &0.87&0.57&1.0 \\
$\sigma$&0.12 &0.20&0.20&0.22 \\
$\log{(M_1/M_{\odot})}$&11.24 &11.27&11.24&11.36\\
\hline
$A$& 0.238$\pm$0.044    &0.219$\pm$0.022   &   0.202$\pm$0.021 &0.172$\pm$0.015 \\
$B$ & -1.947$\pm$0.509&-1.706$\pm$0.263  & -1.493$\pm$0.241 &-1.141$\pm$0.178 \\
\hline
\end{tabular}\vspace{0.3cm}
   \caption{\textit{Top:} Parameters used in Eq.\,\ref{eq:comp_leauthaud} from \citet[]{2016MNRAS.457.4021L} to correct for stellar-mass incompleteness. \textit{Bottom:} Parameters of the linear fits $\rm{\log{(R_{DECaLS}/kpc)}}=$$\,A\,\rm{\log{(M_{\star}/M_{\odot})}}$$\,+\,B$ to the stellar mass-size relations shown in Figure \ref{fig:masssize}.}
 \label{tab:param_mass_size}
 \end{table*}
Figure \ref{fig:masssize} displays the circularized effective radius as a function of stellar mass for the DECaLS LOWZ (upper row) and CMASS (lower row) samples, respectively, in four bins of redshift ($0.2\leq z<0.3$ and $0.3\leq z<0.4$ for LOWZ; $0.43\leq z<0.55$ and $0.55\leq z<0.6$ for CMASS). The density contours are approximately corrected from stellar-mass incompleteness using the analytic formula from \citet[]{2016MNRAS.457.4021L}:
\begin{equation}
c=\frac{f}{2} \left[1+\rm{erf}{\left(\frac{\log{M_{\star}/M_1}}{\sigma}\right)} \right],
\label{eq:comp_leauthaud}
\end{equation}
where the parameter values are chosen at the mean redshift of our samples, see Table\,\ref{tab:param_mass_size}.
As expected, we find a correlation, although mild, between effective radius and stellar mass in our cross-matched BOSS-DECaLS samples. The mean size estimates in bins of stellar mass are displayed on top of each distribution as blue points; the error bars correspond to the $\pm1\,\sigma$ dispersion around the mean. 
A linear fit of the form $\rm{\log{(R_{DECaLS}/kpc)}}=\,\textit{A}\,\rm{\log{(M_{\star}/M_{\odot})}}\,+\,\textit{B}$ is also shown in each panel of Figure \ref{fig:masssize} as a blue solid line; the corresponding parameters are given in Table\,\ref{tab:param_mass_size}. The slope of the mass-size relation increases mildly across our redshift range $0.2\leq z<0.7$, with values of $A\sim0.17-0.24$. \\\\
\indent BOSS provides unprecedented statistics at the high-mass end, as compared to previous surveys and samples at similar redshifts. Establishing a fair comparison at these stellar masses is therefore tricky. Instead, in Figure \ref{fig:masssize}, we show results from two relatively large lower-mass samples. The first one is a selection of quiescent galaxies observed in CFHT SDSS Stripe 82 \citep[][]{2017MNRAS.469.4523C}, with stellar masses from the S82 Massive Galaxy Catalog\footnote{\url{http://www.ucolick.org/~kbundy/massivegalaxies/}} \citep[S82-MGC;][]{2015ApJS..221...15B}. The second one is composed by early-type galaxies detected using COSMOS \citep[]{2013MNRAS.428.1715H}.
When combined, the BOSS mass-size relation appears as a natural higher-mass continuation of those lower-mass relations, but displaying a significantly 
flatter slope (the typical slope at lower-masses is $A\sim0.47-0.61$). 

The apparent flattening observed in the mass-size relation might be due to residual incompleteness and selection effects that we could not take into account in the analysis, and to the CFHT size calibration. In Figure\,\ref{fig:masssize}, we overplot the linear fit to the uncalibrated relation (black dotted line), which is flatter ($A\sim0.20-0.45$) than the lower-mass measurements, but steeper than the corrected relation, especially towards higher redshifts. By comparing these two fits, one can appreciate the effect of the CFHT calibration on the DECaLS size estimates, which are reduced by a factor $\sim0.5-0.25$\,dex.
Note also that the size correction has a stronger effect on the higher redshift bins (i.e., CMASS), as expected from the right panel of Figure\,\ref{fig:corr}. 

The possibility remains that the apparent flattening of the mass-size relation towards the high-mass end is related to the well-documented curvature of scaling relations for early-type galaxies \citep[see e.g.,][]{2007MNRAS.377..402D, 2009MNRAS.394.1978H, 2011MNRAS.412L...6B, 2013ApJ...769L...5K, 2013MNRAS.432.1862C, 2013MNRAS.432.1709C, AMD2016a, AMD2016b, 2017MNRAS.468...47M}. In BOSS, particularly, this phenomenon was reported by \citet[][]{AMD2016b} when analysing the intrinsic $L-\sigma$ relation for the red sequence population. In Section\,\ref{sec:comparison}, we discuss possible interpretations of this result.

\subsection{The redshift-size relation of LRGs at $0.2<z<0.7$}
\label{sec:z-size}
We have analysed the redshift evolution of the average size of massive LRGs from the BOSS-DECaLS cross-matched samples. This measurement, due to the mass-size relation itself, is very sensitive to the particular stellar mass range observed, so comparisons with previous results should be taken with caution.

\begin{figure}
\begin{center}
\includegraphics[width=1.05\linewidth]{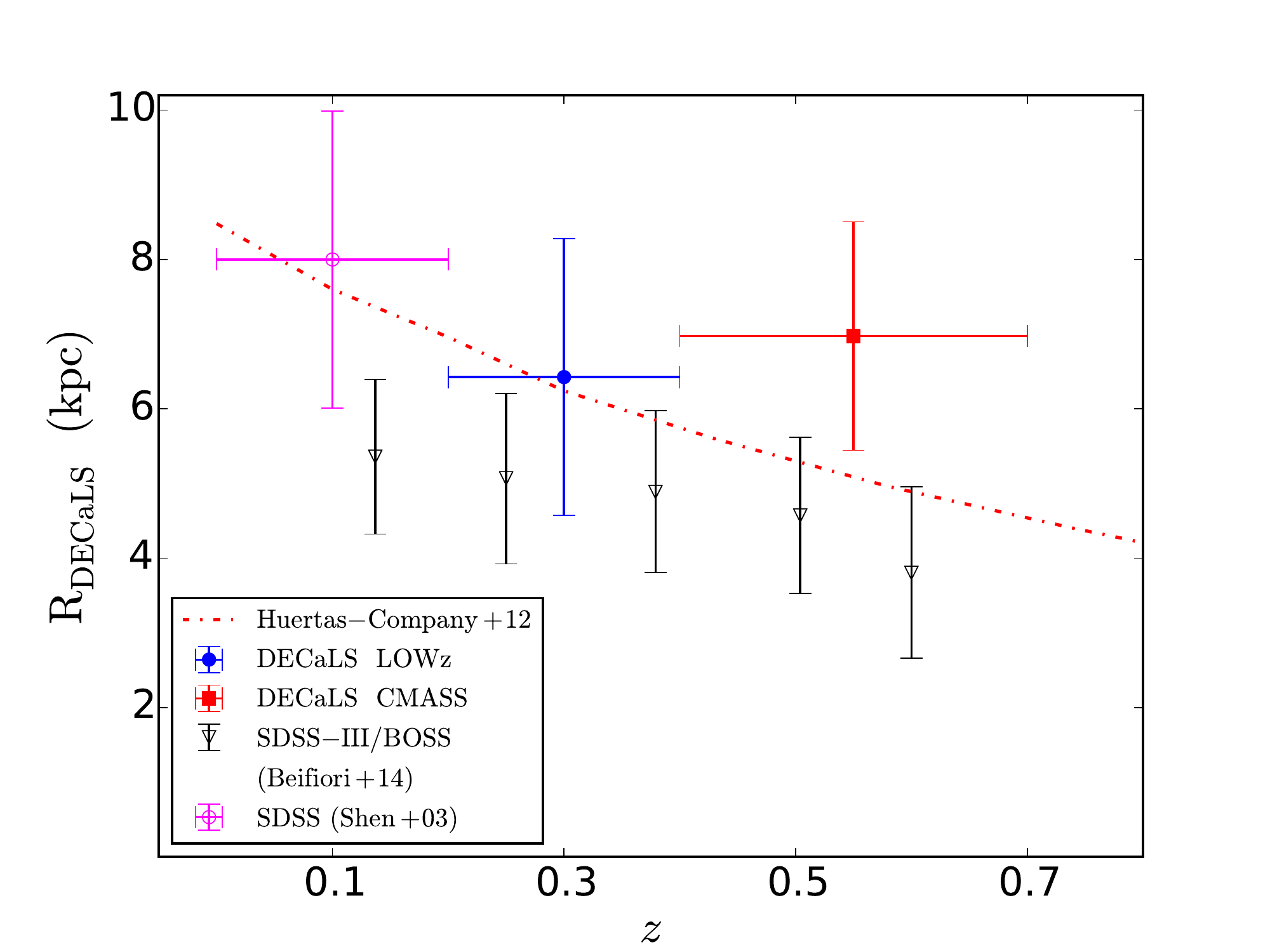}
\caption{Redshift-size relation of our DECaLS CMASS (red filled square) and LOWZ (blue filled point) galaxies, compared to the SDSS-III/BOSS (black empty triangles) results from \citet[]{2014ApJ...789...92B}. We also show the $z\sim0.1$ SDSS Main galaxy sample measurement from \citet[]{2003MNRAS.343..978S} (magenta empty point) which, combined with our results, suggests a mildly declining redshift trend. The dot-dashed line is the fit to the COSMOS ETGs with $11.2< \log{(\rm{M_{\star}/M_{\odot}})} <12$ \citep[]{2013MNRAS.428.1715H}.}
\label{fig:zsize}
\end{center}
\end{figure}

Figure \ref{fig:zsize} displays the mean effective radius of our LOWZ (blue point) and CMASS (red square) samples, in which only galaxies with De Vaucouleurs profile are considered; the error bars correspond to $\pm1\,\sigma$ scatter around the mean. Our results are obtained by integrating over the entire stellar mass range. The empty black triangles represent previous estimates from SDSS and SDSS-III/BOSS \citep[]{2014ApJ...789...92B} calibrated against HST/COSMOS data and selected in a narrow bin of stellar mass. 

The redshift evolution of the DECaLS early-type galaxy sizes calibrated using CFHT data is overall consistent with a flat trend, i.e. no evolution. This is in good agreement with CFHT observations in Stripe 82 of quiescent ETGs \citep[][]{2017MNRAS.469.4523C}. However, when we combine our nearly flat results with the SDSS measurements at $z\sim 0.1$ \citep[][empty magenta point]{2003MNRAS.343..978S}, the evolutionary trend mildly declines with redshift and reconciles with \citet[]{2014ApJ...789...92B}. The effective radius estimates presented by \citet[]{2014ApJ...789...92B} are systematically smaller than our results and their evolutionary trend is overall similarly flat.

Interestingly, when we limit our measurements to very high masses, $\log{(\rm{M_{\star}/ M_{\odot}})}>11.8$, we find a slope steeply declining with redshift. This is in line with current estimates for very massive ETGs in ULTRAVISTA and CANDELS/3D-HST \citep[]{2017ApJ...837..147H} and with the massive ETGs at $11.2< \log{(\rm{M_{\star}/M_{\odot}})} <12$ observed in COSMOS \citep[]{2013MNRAS.428.1715H}.  

\section{Comparison with previous studies}
\label{sec:comparison}

We have measured the stellar mass-size relation for massive early-type galaxies within the redshift range $0.2< z < 0.7$.  When compared with lower-mass results, our measurement shows a relative flattening of this relation, especially at higher redshift.

At face value, it seems that the observed flattening of the mass-size relation could be related to the well-documented curvature of the scaling relations towards the high-mass end, which has been extensively addressed in the literature for early-type galaxies \citep[][]{2009MNRAS.394.1978H, 2007MNRAS.377..402D, 2011MNRAS.412L...6B, 2013ApJ...769L...5K, 2013MNRAS.432.1862C, 2013MNRAS.432.1709C, AMD2016b}. 
In particular, \citet[]{2009MNRAS.394.1978H} studied the stellar mass-size relation in a sample of $\sim50,000$ SDSS ETGs at $z\sim0.1$ and found evidence for a deviation from the linear behaviour: galaxies with $\rm{\log{(M_{\star}/M_{\odot})}}\gtrsim11.5$ have larger sizes than expected. The slope of the regression line depends on the weighting scheme adopted to correct from survey incompleteness and ranges from $A\sim1$ (unitary weights) to $A\sim0.47$ ($1/V_{\rm{max}}(L)$ weights). 
\citet[]{2011MNRAS.412L...6B} demonstrated that different scaling relations for ETGs all point to two preferential mass scales, $3\times10^{10}$ and $2\times10^{11}\,\rm{M_{\odot}}$, as places where fundamental physical processes happen. 
\citet[][]{2013ApJ...769L...5K} investigated the Faber-Jackson correlation between velocity dispersion $\sigma$ and total galaxy luminosity separately
for elliptical galaxies with and without cores. Using the mass-to-light ratio, they related $\sigma$ to the stellar mass. They found that the velocity dispersion of core ellipticals increases much more slowly with luminosity and mass, compared to the coreless galaxies. They claimed that this is an evidence for dry major mergers as the dominant growth mode of the most massive elliptical galaxies. 
\citet[]{AMD2016b} found a steep slope and small scatter for the L-$\sigma$ relation of the massive red sequence population at $z\sim0.55$ using the CMASS sample. 

Although our measurement, in combination with lower-mass results, seems generally consistent with the curvature of the scaling relations towards the high-mass end, it is noteworthy that this behaviour appears to go in the opposite direction to what is reported by \citet[]{2009MNRAS.394.1978H} at low redshift. As mentioned above, they find that SDSS ETGs at the high-mass end are progressively larger than expected (from a linear relation). Establishing a fair comparison is, however, hindered by sample differences. 
Besides focusing on a different redshift range, their conclusion is drawn mostly from an intermediate-mass sample (the high-mass end corresponds to the tail of the distribution), whereas our results are obtained from a larger sample covering exclusively the high-mass end (and after comparing with independent lower-mass measurements at the same redshift). Follow-up work will be specifically devoted to addressing this question.

We have also measured the redshift evolution of the average size of massive early-type galaxies from $z=0.7$. Our results are consistent with a non-evolving scenario. This conclusion is in agreement with results from \citet[]{2017ApJ...851...34B}, who detected no growth in the stellar mass of massive (i.e., $\rm{\log(M_{\star}/M_{\odot})>11.2}$) galaxies over $0.3<z<0.65$. \citet[]{AMD2016b} also found results generally consistent with no evolution of the high-mass end of the L-$\sigma$ relation all the way to $z=0$.


\section{Summary and future work}
\label{sec:summary}

We have studied the morphology, the stellar mass-size relation and the size evolution of the SDSS-III/BOSS DR12 CMASS and LOWZ spectroscopic galaxy samples cross-matched with the DECaLS DR3 ($g,\,r,\,z$) deeper and higher-quality image photometry. The resulting CMASS and LOWZ selections include about 31\% and 23\% of the original BOSS samples. 
We find that the large majority of both populations is composed of early-type galaxies with De Vaucouleurs profiles, while only less than 20\% of them are late-type spirals with exponential profiles. The fraction of ETG clearly increases from LOWZ to CMASS. We calibrate the DECaLS sizes of these galaxies against the available observations from CFHT/Megacam and COSMOS with better image quality. We obtain an excellent agreement between these two independent  corrections and our results are fully consistent with \citet[][]{2011MNRAS.418.1055M} using ZEST \citep[][]{2007ApJS..172..406S} data. By cross-matching our CMASS and LOWZ galaxies with De Vaucouleurs profiles with the Portsmouth \citep[]{2013MNRAS.435.2764M} stellar mass catalog for SDSS-III/BOSS LRGs at $0.2<z<0.7$, we are able to study the high-mass end of the distribution up to $\log{(\rm{M\star/M_{\odot}})}\sim\,12.2$ with unprecedented statistics for 313,026 galaxies over 4380\,deg$^{2}$. Our main results can be summarized as:

\begin{enumerate}
\item the BOSS-DECaLS mass-size relation for massive early-type galaxies exhibits a clear correlation with an apparent flattening in the slope compared to previous estimates from ETGs in CFHT SDSS Stripe 82 at lower masses \citep{2013MNRAS.428.1715H, 2017MNRAS.469.4523C}. Further analysis is needed to determine what causes this behaviour. The apparent flattening might be explained by the fact that scaling relations for the most massive early-type galaxies can be systematically different from the same relations at lower masses \citep[e.g.,][]{AMD2016b, 2009MNRAS.394.1978H, 2011MNRAS.412L...6B, 2013ApJ...769L...5K}. \\

\item we find no evolution in the BOSS-DECaLS ETG sizes over $0.2<z<0.7$. This result is consistent with the non-evolving scenario found by \citet[]{AMD2016b} in the high-mass end of the L-$\sigma$ relation all the way to $z=0$. In addition, it is consolidated by the no-growth detection in the stellar mass of Stripe 82 Massive galaxies within $0.3<z<0.65$ \citep[]{2017ApJ...851...34B}.
If we focus only on the most massive galaxies at $\log{(\rm{M_{\star}/ M_{\odot}})}>11.8$, the slope of their evolution changes to steeply declining with redshift. This is in agreement with current estimates for very massive ETGs in ULTRAVISTA and CANDELS/3D-HST \citep[]{2017ApJ...837..147H} and in COSMOS \citep[]{2013MNRAS.428.1715H}.\\

\item combining our BOSS-DECaLS size measurements with the SDSS results at $z\sim 0.1$ \citep[]{2003MNRAS.343..978S}, the evolutionary trend mildly declines with redshift and reconciles with \citet[]{2014ApJ...789...92B}. This is consistent with a passive evolution scenario for LRGs from $z\sim0.55$ \citep[][]{2013MNRAS.435.2764M, AMD2016a, AMD2016b, 2017ApJ...851...34B}.
\end{enumerate}

This work provides a galaxy sample with unprecedented statistics that can be used to further investigate morphological and size-related aspects in the evolution of LRGs. In addition, this cross-matched sample can be used to study the dependence of clustering on morphological and size-related properties of LRGs. Our cross-matched BOSS-DECaLS CMASS and LOWZ samples with CFHT calibrated sizes are made publicly available for the community on the \textsc{Skies and Universes}\footnote{\url{http://www.skiesanduniverses.org/}} database.

In a follow-up study, we will attempt to deconvolve the uncertainties on the effective radius and the residual incompleteness effects present in the mass-size relation using a similar forward-modeling Bayesian method as the one presented in \citet[]{AMD2016a, AMD2016b}. Within this framework, we will be able to measure the mass-size relation for the {\it{intrinsic}} red sequence population photometrically identified in \citet[]{AMD2016a}. 
We also plan to look at the dust properties and star formation history of these galaxies by cross-matching them with the available data from the infra-red Herschel\footnote{\url{http://sci.esa.int/herschel/}} ESA mission. 

In the near future, the Subaru HSC-CCP\footnote{\url{http://hsc.mtk.nao.ac.jp/ssp/}} Collaboration will provide ulta-deep multicolor images down to $r_{\rm{AB}}\sim28$ with 0.6" median seeing, which wil be key to improve the current constraints on galaxy size and morphology. New-generation spectroscopic surveys such as SDSS-IV/eBOSS \citep[]{2016AJ....151...44D}, DESI \citep[][]{2015AAS...22533607S} and Euclid \citep[]{2011arXiv1110.3193L, 2015arXiv150502165S} will produce enormous data sets with high-resolution out to redshift $z\sim2$. These observations will allow us to better understand the galaxy formation paradigm on small scales, and to coherently link it to the evolution of the large scale structure of our Universe.


\section*{Acknowledgments}

GF is supported by a European Space Agency (ESA) Research Fellowship at the European Space Astronomy Centre (ESAC), in Madrid, Spain. 

AMD acknowledges support from the Funda\c{c}\~ao de Amparo \`a Pesquisa do Estado de S\~ao Paulo (FAPESP), through the grant 2016/23567--4.

GF and FP acknowledge financial support from MINECO grant AYA2014-60641-C2-1-P.

GF and FP are thankful to the Lawrence Berkeley National Laboratory for hosting the first phase of this work. GF and coauthors wish to thank D. Lang for insightful discussions on DECaLS morphology details.

Funding for SDSS-III has been provided by the Alfred P. Sloan Foundation, the Participating Institutions, the National Science Foundation, and the U.S. Department of Energy Office of Science. The SDSS-III web site is \url{http://www.sdss3.org/}.

SDSS-III is managed by the Astrophysical Research Consortium for the Participating Institutions of the SDSS-III Collaboration including the University of Arizona, the Brazilian Participation Group, Brookhaven National Laboratory, Carnegie Mellon University, University of Florida, the French Participation Group, the German Participation Group, Harvard University, the Instituto de Astrofisica de Canarias, the Michigan State/Notre Dame/JINA Participation Group, Johns Hopkins University, Lawrence Berkeley National Laboratory, Max Planck Institute for Astrophysics, Max Planck Institute for Extraterrestrial Physics, New Mexico State University, New York University, Ohio State University, Pennsylvania State University, University of Portsmouth, Princeton University, the Spanish Participation Group, University of Tokyo, University of Utah, Vanderbilt University, University of Virginia, University of Washington, and Yale University. 

The Legacy Surveys consist of three individual and complementary projects: the Dark Energy Camera Legacy Survey (DECaLS; NOAO Proposal ID \# 2014B-0404; PIs: David Schlegel and Arjun Dey), the Beijing-Arizona Sky Survey (BASS; NOAO Proposal ID \# 2015A-0801; PIs: Zhou Xu and Xiaohui Fan), and the Mayall z-band Legacy Survey (MzLS; NOAO Proposal ID \# 2016A-0453; PI: Arjun Dey). DECaLS, BASS and MzLS together include data obtained, respectively, at the Blanco telescope, Cerro Tololo Inter-American Observatory, National Optical Astronomy Observatory (NOAO); the Bok telescope, Steward Observatory, University of Arizona; and the Mayall telescope, Kitt Peak National Observatory, NOAO. The Legacy Surveys project is honored to be permitted to conduct astronomical research on Iolkam Du'ag (Kitt Peak), a mountain with particular significance to the Tohono O'odham Nation.

NOAO is operated by the Association of Universities for Research in Astronomy (AURA) under a cooperative agreement with the National Science Foundation.

This project used data obtained with the Dark Energy Camera (DECam), which was constructed by the Dark Energy Survey (DES) collaboration. Funding for the DES Projects has been provided by the U.S. Department of Energy, the U.S. National Science Foundation, the Ministry of Science and Education of Spain, the Science and Technology Facilities Council of the United Kingdom, the Higher Education Funding Council for England, the National Center for Supercomputing Applications at the University of Illinois at Urbana-Champaign, the Kavli Institute of Cosmological Physics at the University of Chicago, Center for Cosmology and Astro-Particle Physics at the Ohio State University, the Mitchell Institute for Fundamental Physics and Astronomy at Texas A\&M University, Financiadora de Estudos e Projetos, Fundacao Carlos Chagas Filho de Amparo, Financiadora de Estudos e Projetos, Fundacao Carlos Chagas Filho de Amparo a Pesquisa do Estado do Rio de Janeiro, Conselho Nacional de Desenvolvimento Cientifico e Tecnologico and the Ministerio da Ciencia, Tecnologia e Inovacao, the Deutsche Forschungsgemeinschaft and the Collaborating Institutions in the Dark Energy Survey. The Collaborating Institutions are Argonne National Laboratory, the University of California at Santa Cruz, the University of Cambridge, Centro de Investigaciones Energeticas, Medioambientales y Tecnologicas-Madrid, the University of Chicago, University College London, the DES-Brazil Consortium, the University of Edinburgh, the Eidgenossische Technische Hochschule (ETH) Zurich, Fermi National Accelerator Laboratory, the University of Illinois at Urbana-Champaign, the Institut de Ciencies de l'Espai (IEEC/CSIC), the Institut de Fisica d'Altes Energies, Lawrence Berkeley National Laboratory, the Ludwig-Maximilians Universitat Munchen and the associated Excellence Cluster Universe, the University of Michigan, the National Optical Astronomy Observatory, the University of Nottingham, the Ohio State University, the University of Pennsylvania, the University of Portsmouth, SLAC National Accelerator Laboratory, Stanford University, the University of Sussex, and Texas A\&M University.

BASS is a key project of the Telescope Access Program (TAP), which has been funded by the National Astronomical Observatories of China, the Chinese Academy of Sciences (the Strategic Priority Research Program "The Emergence of Cosmological Structures" Grant \# XDB09000000), and the Special Fund for Astronomy from the Ministry of Finance. The BASS is also supported by the External Cooperation Program of Chinese Academy of Sciences (Grant \# 114A11KYSB20160057), and Chinese National Natural Science Foundation (Grant \# 11433005).

The Legacy Survey team makes use of data products from the Near-Earth Object Wide-field Infrared Survey Explorer (NEOWISE), which is a project of the Jet Propulsion Laboratory/California Institute of Technology. NEOWISE is funded by the National Aeronautics and Space Administration.

The Legacy Surveys imaging of the DESI footprint is supported by the Director, Office of Science, Office of High Energy Physics of the U.S. Department of Energy under Contract No. DE-AC02-05CH1123, by the National Energy Research Scientific Computing Center, a DOE Office of Science User Facility under the same contract; and by the U.S. National Science Foundation, Division of Astronomical Sciences under Contract No. AST-0950945 to NOAO.

\bibliographystyle{mnras}
\bibliography{./bibliography}

\end{document}